\documentclass[showpacs,preprint,nofootinbib,superscriptaddress,citeautoscript,eqsecnum,12pt]{article}
\usepackage{latexsym}
\usepackage{amsmath}
\usepackage{amsfonts}
\usepackage{amssymb}
\usepackage{amscd}
\usepackage{bbm}
\usepackage{fancybox}
\usepackage{cite}
\usepackage{amsmath,amsfonts,amsbsy}
\usepackage{pstricks,pst-node}
\usepackage[small,bf,hang]{caption2}
\usepackage{slashed}
\usepackage{graphicx}
\usepackage{epsfig}
\usepackage{psfrag}
\usepackage{comment}
\usepackage{hyperref}

\usepackage[utf8x]{inputenc}
\usepackage{relsize}
\usepackage{dcolumn}

\usepackage{float}

\psset{unit=1.3cm,linewidth=.5pt,radius=.2}  % controls the size of diagrams

\usepackage{multirow}                     % tables cells spanning multiple rows
\usepackage{float}                          % to force floaters to stay Here
\usepackage{lscape}                         % landscape laid out pages
\usepackage{bm}

\newcommand{\cD}{{\cal D}}

\newcommand{\cG}{{\cal G}}

\newcommand{\cL}{{\cal L}}
\newcommand{\cM}{{\cal M}}

\newcommand{\cP}{{\cal P}}

\newcommand{\beq}{\begin{equation}}
\newcommand{\eeq}{\end{equation}}
\newcommand{\bi}{\begin{itemize}}
\newcommand{\ei}{\end{itemize}}

\makeatletter
\newcommand{\eqnum}{\refstepcounter{equation}\textup{\tagform@{\theequation}}}
\makeatother
\newcommand{\bt}{\begin{tabular}}
\newcommand{\et}{\end{tabular}}
\newcommand{\bc}{\begin{center}}
\newcommand{\ec}{\end{center}}

\def\one{{\hbox{ 1\kern-.8mm l}}}
\newcommand{\Dslash}{\not{\hbox{\kern-4pt $D$}}}
\newcommand{\pdslash}{\not{\hbox{\kern-2pt $\partial$}}}
\newcommand{\nn}{\nonumber}

\newcommand{\be}{\begin{equation}}
\newcommand{\ee}{\end{equation}}
\newcommand{\bea}{\begin{eqnarray}}
\newcommand{\eea}{\end{eqnarray}}
\newcommand{\ba}{\begin{array}}
\newcommand{\ea}{\end{array}}

\def\bbox{{\,\lower0.9pt\vbox{\hrule \hbox{\vrule height 0.2 cm
\hskip 0.2 cm \vrule height 0.2 cm}\hrule}\,}}
\newcommand{\dsl}{\pa \kern-0.5em /}

\begin{document}

\begin{titlepage}%1
\begin{center}

\hfill UG-2014-93

\vskip 1.5cm

{\Large \bf Extended massive gravity in three dimensions}

\vskip 1cm

{\bf Hamid R.~Afshar, Eric A.~Bergshoeff and Wout Merbis}

\vskip 25pt

{\em Centre for Theoretical Physics,
University of Groningen, \\ Nijenborgh 4, 9747 AG Groningen, The
Netherlands \vskip 5pt }

{email: {\tt h.r.afshar@rug.nl, E.A.Bergshoeff@rug.nl,  w.merbis@rug.nl}} \\
\vskip 10pt

\end{center}

\vskip 1.5cm

\begin{center} {\bf ABSTRACT}\\[3ex]
\end{center}
Using a first order Chern-Simons-like formulation of gravity we systematically construct higher-derivative extensions of general relativity in three dimensions. The construction ensures that  the resulting
higher-derivative gravity theories are free of scalar ghosts. We canonically analyze these theories and construct the gauge generators and the boundary central charges. The models we construct are all consistent with a holographic $c$-theorem which, however, does not imply that they are unitary. We find that Born-Infeld gravity in three dimensions is contained within these models as a subclass.

\end{titlepage}

\newpage
\tableofcontents

\section{Introduction}
Three-dimensional gravity models are an interesting playground to study problems in quantum gravity. In particular, three dimensional general relativity (GR) with or without cosmological constant is known to be described by a Chern-Simons (CS) gauge theory \cite{Achucarro:1986vz,Witten:1988hc}, at least classically. 
The presence of a negative cosmological constant not only makes it possible to have black hole solutions \cite{Banados:1992wn,Banados:1992gq} but also leads to the boundary global degrees of freedom by defining an asymptotic boundary and imposing appropriate boundary conditions \cite{Brown:1986nw}.
In spacetimes with asymptotic boundaries, one can define the asymptotic symmetry group  as the group of boundary condition preserving gauge transformations. For asymptotically locally anti-de~Sitter (AdS$_3$) spacetimes, the asymptotic symmetry group consists of 
two copies of the Virasoro algebra with a classical central extension. This approach to quantum gravity is  one  of the earliest applications of the AdS/CFT correspondence which organizes our understanding of a quantum gravity theory in terms of a dual conformal field theory (CFT) and vice versa. 

Although the lack of local degrees of freedom in three dimensional GR can be regarded as a technical simplification, it makes it less interesting from a perturbative field theoretic point of view in which propagating degrees of freedom play an important role. 
One way to compensate for this shortcoming in a purely gravitational manner is to add higher-derivative terms to GR, which can lead to new massive spin-2 modes in the spectrum \cite{Deser:1982wh,Bergshoeff:2009hq}. From the CFT point of view this would correspond to deforming the dual CFT by a new operator which couples to this new massive bulk mode. Deformations in unitary two dimensional CFTs are restricted by Zamolodchikov's $c$-theorem \cite{Zamolodchikov:1986gt}. The presence of a holographic $c$-theorem \cite{Freedman:1999gp} can clarify the role of the higher-curvature terms \cite{Myers:2010tj}, as it implies a restriction on the coupling constants of these higher-derivative interactions in the bulk. 

The main aim of this paper is to systematically construct higher-derivative extensions of three dimensional gravity which are free of scalar ghosts and consistent with a holographic $c$-theorem. 
In this construction, we exploit a first order formulation which is denoted as the `Chern-Simons-like' formulation \cite{Hohm:2012vh,Bergshoeff:2014bia}. In this formulation  
the dynamical fields are a set of one-forms, denoted by flavor indices $r,s,t,\ldots$, taking values in the three-dimensional Lorentz group SO(2,1),
\begin{equation}
a^r = (a^{r\,a}_{\mu} dx^{\mu}) J_a\,.
\end{equation}
The CS-like Lagrangian three-form constructible from these Lorentz-valued one-forms can be defined as
\begin{equation}\label{CSlike}
L_{\text{\tiny{CS-like}}} =  \frac12\left\langle  g_{rs} a^r \wedge da^s + \tfrac23 f_{rst} a^r \wedge a^s \wedge a^t \right\rangle\,.
\end{equation}
Here $g_{rs}$ is a symmetric and invertible metric on the flavor space, and the coupling constants $f_{rst}$ define a totally symmetric flavor tensor.
This construction is completely gauge invariant under SO(2,1) once we use the spin-connection $\omega$ as the gauge field and the trace over Lorentz indices in the three dimensional representation of SO(2,1) \cite{Witten:2007kt}.
% They are defined  way in terms of wedge products of one-form fields in three dimensions 
The corresponding bilinear form, structure constants and covariant derivative are given by
\bea
\langle J_a,\,J_b\rangle=\eta_{ab}\,,\qquad[J_a,J_b]=\epsilon_{ab}{}^cJ_c \qquad\text{and}\qquad \mathcal{D}\equiv d+\tfrac{1}{2}[\omega,\,\,\,]\,.
\eea
The dualized curvature two-form is then given by\,\footnote{Unless stated explicitly, we normally use the notation in which wedge products are implicit.}
\bea
R^a=\mathcal{D}\omega^a=d\omega^a+\frac{1}{2}\epsilon^a{}_{bc}\omega^b \omega^c\,.
\eea

 Whenever the combinations $\epsilon^{a}{}_{bc} f^{r}{}_{st}$ are the structure constants of some Lie algebra and $g_{rs}\eta_{ab}$ a bilinear form on this algebra, then the theory defined by \eqref{CSlike} is actually a Chern-Simons gauge theory. There are two parity preserving\footnote{Here, by `parity preserving' theories we mean those Lagrangians which have a definite parity, $PL=\pm L$, while `parity violating' Lagrangians have no definite parity.} gravity models in three dimensions for which this is the case. They are Einstein gravity, of even parity, and conformal gravity\footnote{This model is sometimes referred to as conformal Chern-Simons gravity, denoted as CSG \cite{Afshar:2011yh,Afshar:2011qw}.} which is odd under parity. They have the following first order actions:
\begin{align}
&\text{Parity-even},\qquad S_{\text{\tiny{Einstein}}}\equiv S_0 = - \frac{1}{\kappa^2} \int \left\langle e\wedge\left( R - \tfrac{\Lambda_0}{3}  e\wedge e\right)\right\rangle\,, \label{EHG}\\
&\text{Parity-odd},\qquad S_{\text{\tiny{Conformal}}}\equiv S_1=\frac{1}{2\kappa^{2}\mu} \int \left\langle \omega\wedge \left(d \omega + \tfrac{2}{3}\omega\wedge  \omega\right)+ 2f\wedge \mathcal De\right\rangle\,. \label{CSG}
\end{align}

Here $\kappa^{2} = 8 \pi G$ is the three dimensional Planck mass and $\Lambda_0$  the cosmological constant with the dimension of (mass)$^{2}$ while $\mu$ is a parameter with  the dimension of mass. This amounts to a dimensionless coupling constant for conformal gravity which is a conformally invariant theory.
Due to the lack of any local degrees of freedom, these models can be written purely as Chern-Simons gauge theories for SO(2,2) and SO(2,3) respectively, where $e$, $\omega$ and $f$ correspond to the gauge fields for translations, rotations and special conformal transformations, respectively \cite{Achucarro:1986vz,Horne:1988jf,Witten:1988hc}. For a
recent discussion of these theories, see  \cite{Afshar:2013bla}. 

% All fields in the models \eqref{EHG} and \eqref{CSG} are one-forms with values in the three dimensional Lorentz gauge group
%\bea
%e=(e_\mu{}^a dx^\mu)J_a\,,
%\qquad\omega=(\omega_\mu{}^a dx^\mu)J_a\,\quad \text{and}\quad f=(f_\mu{}^a dx^\mu)J_a\,.
%\eea
%The first order formulation of Einstein and conformal gravities \eqref{EHG} and \eqref{CSG} in spite of having a CS gauge theory formulation, fall into a broader class of first order actions denoted as CS-like models \cite{Hohm:2012vh}; they look like CS gauge theories as they are SO(2,1) gauge theories with the spin-connection $\omega$ as the gauge field \cite{Witten:2007kt}. They are defined completely in a gauge invariant way in terms of wedge products of Lorentz-valued one-form fields in three dimensions with the trace over Lorentz indices in the three dimensional representation of SO(2,1).

%
%The appearance of the necessary secondary second-class constraints needed to remove additional ghost-like scalar degrees of freedom follows from the particular way in which we construct the models.

%For all other cases the theory defined by \eqref{CSlike} is only CS-like, but some of the desirable properties of CS gauge theories are maintained. In particular, the Hamiltonian form is relatively simple \cite{Hohm:2012vh,Bergshoeff:2014bia}. 
In this work we will consider extensions of the above theories in the CS-like formulation to include dynamical spin-2 degrees of freedom by introducing sufficiently many auxiliary one-forms in a parity preserving way. After integrating out these auxiliary one-forms, the resulting theory is a parity-even or a parity-odd higher-derivative theory of gravity. The most general set of parity violating models can be constructed by combining the parity-even and odd theories. The first of these parity violating models is topologically massive gravity (TMG) \cite{Deser:1982wh}, which is the sum of the actions \eqref{EHG} and \eqref{CSG} and propagates a single massive helicity-2 state with one local degree of freedom. An example of a  parity preserving extension which describes two helicity-$\pm 2$ states, with two degrees of freedom, is `new massive gravity' (NMG) \cite{Bergshoeff:2009hq,Bergshoeff:2009aq}.

Both TMG and NMG  may be described in terms of a first order CS-like formulation  \cite{Carlip:2008qh,Blagojevic:2008bn,Blagojevic:2010ir,Hohm:2012vh}. The first order formulation of conformal gravity given in \eqref{CSG}, which is a three-derivative action in terms of the metric is an example of how one can exchange a higher-derivative action for a first order action containing auxiliary fields. In the case of conformal gravity this auxiliary field is the one-form  $f^a$ \cite{Afshar:2011yh,Afshar:2011qw} --- see \cite{Afshar:2014rwa} for a recent review.
In \cite{Blagojevic:2010ir,Hohm:2012vh} this approach was extended to four-derivative actions by introducing two extra auxiliary one-form fields ($f^a$, $h^a$) to obtain NMG, which for future reference we denote by $S_2$:
\begin{equation}\label{NMG}
S_{\text{\tiny{NMG}}}\equiv S_2 = S_0 - \frac{1}{\kappa^2m^2} \int \left\langle f\wedge \left(R + e\wedge f \right)+h\wedge \cD e   \right\rangle\,.
\end{equation}
The equivalent four derivative action is recovered after integrating out the two auxiliary fields.

Originally, NMG was not found in the first order form given above. Instead, it was constructed by  extending GR in 3D with higher-curvature $R_{\mu\nu}R^{\mu\nu}$ and $R^2$ terms. It was found that the theory describes the two degrees of freedom of a massive spin-2 degree of freedom only for a particular combination of higher-curvature terms \cite{Bergshoeff:2009hq,Bergshoeff:2009aq,deRham:2011ca}. Moving away from this special combination introduces a third scalar degree of freedom corresponding to a Boulware-Deser ghost mode \cite{Boulware:1973my}.  Interestingly, it is only possible to write down a CS-like formulation of $R^2$ extended gravity if the higher-curvature terms occur precisely in the ghost-free NMG combination. We will use this observation  as a guiding principle to construct further generalizations of higher-derivative massive gravity in three dimensions which, as we will show explicitly, are free of scalar ghost excitations.  
%Our construction method resembles the way we  obtained the first order formulation of NMG. In the NMG case we introduced  two extra auxiliary fields. Upon solving for these auxiliary fields and  substituting the solutions  back into the Lagrangian we found  the conventional higher-derivative metric formulation of NMG. In this work we extend this reasoning to include more auxiliary fields, which all may be solved for in terms of higher-derivatives acting on the dreibein leading to a higher-derivative extension of NMG. 

For the parity-even sector, we start by considering Einstein gravity \eqref{EHG}. First, we  include two auxiliary fields to obtain NMG. Next, we will show that by adding an additional set of two auxiliary fields we obtain a six derivative  theory which generically propagates {\sl two} massive spin-2 modes. The resulting theory is a combination of the $R^3$ terms considered in \cite{Sinha:2010ai} as extended NMG  and  $R \square R$ terms also considered in \cite{Bergshoeff:2012ev} as parity-even tricritical (PET) gravity.  We analyze the linear spectrum of the  theory and find that even though for a general choice of parameters the  scalar ghosts are absent, one of the two massive spin-2 modes is either tachyonic or a ghost. An exception is cubic extended NMG which only propagates one massive spin-2 mode which can be removed from the linear spectrum via a fine tuning in the couplings of the $R^2$ and $R^3$ terms \cite{Sinha:2010ai}. Furthermore, at special points in the parameter space, either one or both of 
the massive modes become massless and a degeneracy takes place. At these points the dual CFT obtains a non-diagonalizable Jordan cell \cite{Grumiller:2008qz} and after adopting the appropriate logarithmic boundary conditions \cite{Grumiller:2008es,Henneaux:2009pw,Maloney:2009ck} becomes a logarithmic CFT (LCFT) --- see \cite{Grumiller:2013at} for a recent review.

In the parity-odd sector, we start with conformal gravity \eqref{CSG}. In this case, adding two auxiliary fields will break the conformal symmetry of the original theory, and the resulting five derivative  theory propagates {\sl three} local degrees of freedom: a `partially massless' mode \cite{Deser:2001pe} and the two helicity-$\pm2$ states of a massive spin-2 mode. In this case, there is no way to tune the mass of the massive mode to zero. However, there is a special point where the massive mode degenerates with the partially massless mode.

In both sectors, all auxiliary fields and consequently all higher-derivative terms in the action are engendered by the Schouten and Cotton tensors, defined below in \eqref{Schoutenten} and \eqref{scho-cot}. The novelty, due to the presence of the Cotton tensor, is that it allows for actions containing terms with derivatives of curvatures consistent with a holographic $c$-theorem. In this way the class of higher-derivative theories admitting a holographic $c$-theorem in three dimensions is larger than the class of theories considered in \cite{Sinha:2010ai,Paulos:2010ke}, which only included higher-curvature terms containing Schouten tensor.

% Having explicitly constructed the higher-derivative models of massive 3D gravity,  

This paper is organized as follows. In section \ref{Extended} we describe the general procedure to construct higher-derivative gravities in the  CS-like formulation. We derive a general action principle for the parity-even theories and the parity-odd theories separately and show that all of these extensions contain the required secondary (and second-class) constraints needed to remove the Boulware-Deser ghost. In section \ref{ExtendedNMG} we explicitly construct the six  and eight derivative  extension of GR and analyze the linear spectrum of the former to verify that it propagates two massive spin-2 excitations. By explicitly deriving the kinetic and the mass terms of the bulk modes in the Lagrangian, we show that one of the two massive spin-2 modes is either tachyonic or a ghost. There are, however, critical lines and points in the parameter space where the massive modes either disappear or become massless and degenerate with the pure gauge 
mode. In section \ref{Holography} we discuss AdS holography for these models. We identify the conserved boundary charges and show that, when adopting Brown-Henneaux boundary conditions, the asymptotic symmetry algebra consists of two copies of the Virasoro algebra. We compute the semi-classical central charge for the six derivative model and give an expression for the new anomalies when the central charge becomes zero. The dual CFT at these points is expected to be logarithmic as they lead to the appearance of Jordan cells. Furthermore, we show the consistency of all the theories constructed in this way with a holographic $c$-theorem. 
Finally, we have included two appendices. In appendix \ref{can}  we discuss the identification of the first class constraints of the CS-like theories which generate the gauge symmetries and the corresponding boundary charges.
Appendix \ref{ExtendedCSG} is devoted to the analysis of the parity-odd five-derivative extension of the gravitational Chern-Simons term.

\section{Extended massive gravity models}\label{Extended}
In this section, we give a procedure to derive higher-derivative extensions of 3D GR which propagate multiple massive spin-2 particles. The extensions are obtained from an auxiliary field formalism which, as we
 will show, guarantees the freedom from scalar ghosts. However, as we will show in section \ref{linearth},   the higher-derivative nature of the theory does lead to the presence of massive spin-2 ghosts. %We work out the extensions up to eight order in derivatives explicitly.

 Our starting point is a first order, Chern-Simons-like formalism \cite{Hohm:2012vh,Bergshoeff:2014bia}, defined by a Lagrangian three-form depending on the dreibein $e^a$, the dualized spin-connection $\omega^a$ and a number of new auxiliary Lorentz vector valued one-forms $f^a_I$ and $h^a_I$. The advantage of this approach is that it automatically leads to higher-derivative  terms which are free of scalar ghosts, as we will show below.

% \subsection{General Procedure}%and the Absence of Scalar Ghosts

The construction is such that  the field equations will always ensure the vanishing of the torsion two-form
\bea\label{torsion-const}
T^a=\mathcal{D}e^a=de^a+\epsilon^a{}_{bc}\omega^b e^c=0\,.
\eea
Assuming the invertibility of the dreibein it is possible to solve this equation for the spin-connection in terms of the dreibein: $\omega^a = \omega^a (e)$.
Varying the Einstein gravity action \eqref{EHG} w.r.t. $e^a$ gives the equation $R^a = \frac{1}{2}\Lambda_0\epsilon^{abc} e_b e_c$, which can be written in the metric form
as
$ G_{\mu\nu} + \Lambda_0\,g_{\mu\nu} = 0$, where $G_{\mu\nu}=R_{\mu\nu}-\tfrac{1}{2}Rg_{\mu\nu}$ is the Einstein tensor.

Varying the conformal gravity action \eqref{CSG} w.r.t. $f^a$, $\omega^a$ and $e^a$ gives the field equations,
\begin{align}%\label{CSGeom}
& \cD e^a = 0\,, \nonumber \\
& R^a + \epsilon^{a}{}_{bc}\, f^b e^c=0\,,\label{Schouten}\\
& \cD f^a=0 \nonumber \,.
\end{align}
Assuming the invertibility of the dreibein, the auxiliary field $f^a$ can be solved for in terms of the curvature two-form as
\bea\label{Schoutenten}
f_{\mu\nu} \equiv f_{\mu}{}^a e_{\nu\,a} = -\left(R_{\mu\nu}-\tfrac{1}{4}R\,g_{\mu\nu}\right)\equiv - S_{\mu\nu}(e)\,.
\eea
The last equation in \eqref{Schouten} then gives a third order differential equation for the  dreibein: $C_{\mu\nu}(e)\equiv e^{-1} \epsilon_{(\mu|}{}^{\alpha\beta} \nabla_{\alpha} S_{\beta |\nu)}=0$.
Here $S_{\mu\nu}$ and $C_{\mu\nu}$ are the symmetric Schouten and Cotton tensors respectively, constructed from the dreibein $e^a$.

%%%%

% - \frac{1}{\kappa^2} \int \left\langle \sigma e \wedge R - \tfrac{\Lambda_0}{3} e \wedge e \wedge e \right\rangle
% Here $\sigma$ is the NMG sign parameter $\sigma = \pm1$.
Varying the NMG action \eqref{NMG}  the following field equations arise:
\begin{equation}
\begin{split} \label{NMGeom}
& \cD e^a = 0\,,  \\
& R^a + \epsilon^{abc} e_b f_{\,c} = 0 \,, \\
& \cD f{}^a + \epsilon^{abc} e_b h_{\,c} = 0\,, \\
& \cD h^a +\tfrac{1}{2}\epsilon^{abc}\left(f_bf_c - 2  m^2e_bf_c - m^2\Lambda_0 e_be_c\right) = 0\,.
\end{split}
\end{equation}
The first equation in \eqref{NMGeom} is the torsion constraint, the second one is solved as in \eqref{Schoutenten} and the third equation gives,
\begin{equation}\label{scho-cot}
h{}_{\mu\nu} \equiv h{}_{\mu}{}^a e_{\nu\,a} = e^{-1} \epsilon_{(\mu|}{}^{\alpha\beta} \nabla_{\alpha} S_{\beta \,|\nu)}= C_{\mu\nu}(e)\,.
\end{equation}
The last equation in \eqref{NMGeom} then leads to an equation for the dreibein which is fourth order in derivatives.

Looking at equations \eqref{Schouten} and \eqref{NMGeom} suggests to continue this logic to obtain arbitrarily higher-derivative extensions of 3D GR.
Inspired by the above we consider the following schematic  extension of the equations:
\begin{equation}
\begin{split} \label{ENMGeom}
1 \qquad \qquad & \cD e^a = 0\,, \\
2 \qquad \qquad & R^a + \epsilon^{abc} e_b f_{1\,c} = 0 \,, \\
3 \qquad \qquad & \cD f_1{}^a + \epsilon^{abc} e_b h_{1\,c} = 0\,, \\
4 \qquad \qquad & \cD h_1{}^a + \epsilon^{abc} e_b f_{2\,c} + \ldots = 0\,, \\
\vdots \qquad\qquad&  \qquad \quad \vdots \\
2N+1 \qquad & \cD f_N{}^a + \epsilon^{abc} e_b h_{N\,c}+ \ldots = 0\,,\\
2N+2 \qquad & \cD h_N{}^a + \ldots = 0\,.
\end{split}
\end{equation}
The structure of these equations is such that they may be solved one after the other, starting with the first one,  in terms of derivatives acting on the dreibein. The number appearing before each equation denotes the maximum number of derivatives of the dreibein which may appear in the equation after all fields have been solved. The dots denote terms which may contain fewer derivatives or an equal number of derivatives  acting on $e^a$.

The first equation in \eqref{ENMGeom} solves for the spin-connection in terms of the dreibein. The next two equations are already solved as in \eqref{Schoutenten} and \eqref{scho-cot}. The other auxiliary form fields ($f_I$, $h_I$) can be obtained in terms of  $e$ and derivatives acting on it, such that the final equation is a higher-derivative field equation for the dreibein. This set of equations may terminate with an equation for $\cD h_N$ or $\cD f_{N+1}$. The final equation then becomes, an even- or an odd-order partial differential equation for the dreibein corresponding to a parity-even or parity-odd theory respectively.

We can diagramatise the even and odd cases as follows
\begin{align}
\text{Even:}\quad&\mathlarger{\underbrace{\overbrace{\overset{e}{\bullet}\mathlarger\longrightarrow\overset{\omega}{\circ}}^\text{\small{Einstein}}\mathlarger\longrightarrow\overbrace{\overset{f_1}{\circ}\mathlarger\longrightarrow\overset{h_1}{\circ}}^\text{2 dof}}_\text{NMG}\mathlarger\longrightarrow\overset{f_2}{\circ}\cdots\mathlarger\longrightarrow\overset{f_{N}}{\circ}\mathlarger\longrightarrow\overset{h_{N}}{\circ}}\,\,\,\Rightarrow \,2N \,\,\text{dof}\,,N\geq0\nn\\\nn\\
\text{Odd:}\quad&\mathlarger{\overbrace{\overset{e}{\bullet}\mathlarger\longrightarrow\overset{\omega}{\circ}\mathlarger\longrightarrow\overset{f_1}{\circ}}^\text{\small{Conformal}}\mathlarger\longrightarrow\overbrace{\overset{h_1}{\circ}\mathlarger\longrightarrow\overset{f_2}{\circ}}^\text{2 dof}\cdots\mathlarger\longrightarrow\overset{h_{N}}{\circ}\mathlarger\longrightarrow\overset{f_{N+1}}{\circ}}\,\Rightarrow \,2N+1 \,\,\text{dof}\, , N\geq1\nn\\\nn
\end{align}
The sequential form of the diagram shows which fields are solved in terms of which ones. The filled circle denotes the assumption of invertibility of the dreibein. All other fields (open circles) need not be invertible.

\subsection{Action principle}

In both even and odd cases the set of equations \eqref{ENMGeom} can be integrated to an action by the same general procedure. The field with the highest number of derivatives on the dreibein ($h_N$ for even parity, $f_{N+1}$ for odd parity) can be used as a multiplier for the torsion constraint. The field with one derivative less will be used to multiply the second equation, and so on, until half of the field equations have been used. The rest of the field equations then follow from the action by varying the fields with a lower number of derivatives acting on the dreibein. This procedure guarantees that the highest number of derivatives appearing in the action after solving for all the auxiliary fields is $2N+2$ for the parity-even models and $2N+3$ for the parity-odd models. We
find the following actions for the parity-even and parity-odd cases.

\subsubsection*{Parity-even models}\label{sec:2even}

The parity preserving extensions of the  Einstein gravity action \eqref{EHG} in  CS-like form can be obtained from the following recursive action:

\begin{align}\label{SNeven}
S_{2N}=S_{2N-2}+\frac{\kappa^{-2}}{(m^2)^{N}}\int\bigg[\sum_{I+J=N}&\left\langle f_I\wedge\mathcal Dh_{J}\right\rangle+\sum_{I+J+K=N+1}\alpha_{IJK}\left\langle f_I\wedge f_J\wedge f_K\right\rangle\nn\\
+&\sum_{\underset{J,K\neq0}{I+J+K=N}}\beta_{IJK}\left\langle f_I\wedge h_J\wedge h_{K}\right\rangle\bigg]\,,
\end{align}
where $I=0,1,\cdots, N$ in both $f_I$ and $h_I$ with  $(f_0, h_0) \equiv (e,\omega)$.
%and $R=\mathcal{D}\omega$ is the curvature two-form.
The starting value in
this recursive relation is given by the Einstein gravity action $S_0$ given in eq.~\eqref{EHG}. As an example,
the action $S_2$ is already constructed in eq.~\eqref{NMG}.

\subsubsection*{Parity-odd models}\label{sec:2odd}

The parity preserving extension of \eqref{CSG} in CS-like form can be obtained from the following recursive action:

\begin{align}\label{SNodd}
S_{2N+1}=S_{2N-1}+&\,\frac{\kappa^{-2}}{\mu(\mu^2)^{N}}\int\bigg[\sum_{I+J=N}\left\langle h_I\wedge\mathcal Dh_{J}\right\rangle+\sum_{\underset{I,J,K\neq0}{I+J+K=N}}\alpha_{IJK}\left\langle h_I\wedge h_J\wedge h_K\right\rangle\nn\\
+&\sum_{I+J=N+1}\left\langle f_I\wedge\mathcal Df_{J}\right\rangle+\sum_{\underset{K\neq0}{I+J+K=N+1}}\beta_{IJK}\left\langle f_I\wedge f_J\wedge h_{K}\right\rangle\bigg]\,.
\end{align}
% where $\tilde k=\frac{4\pi}{}$ is a dimensionless parameter.
Here $I=0,1,\cdots, N+1$ in $f_I$ and $I=0,1,\cdots, N$ in $h_I$ 
% The indices of $f_I$ in this odd sector run from zero to $N+1$ and of $h_I$ from zero to $N$
with $(f_0,h_0) \equiv (e,\omega)$.  The starting value in
this recursive relation is given by the conformal gravity action $S_1$ given in eq.~\eqref{CSG}. As an example we give here
the explicit form of the next action $S_3$:

\begin{equation}\label{S3}
S_3 = S_1 + \frac{1}{\kappa^2\mu^3} \int \left\langle e\wedge \cD f_2+h_1\wedge \left(R + e\wedge f_1 \right)+\alpha  f_1\wedge \cD f_1   \right\rangle\,.
\end{equation}
We will analyze this model in appendix \ref{ExtendedCSG}.

 Not all couplings $\alpha_{IJK}$ and $\beta_{IJK}$ in \eqref{SNeven} and \eqref{SNodd} are physical. For a given $N$ we have $2N$ auxiliary fields in the even sector and $2N+1$ in the odd case
which can be rescaled such that the same number of coefficients may be set to unity.
In eqs.~\eqref{SNeven}  and \eqref{SNodd} we have already exhausted $N+1$ and $N+2$  of these rescalings, respectively, by canonically normalizing the $N+1$ and $N+2$ kinetic terms.
% \footnote{At each order $N$ the number of  interaction terms is given by the $x^{N}$ coefficient in the Taylor expansion of $\frac{x (x+1)}{(1-x)^3 \left(x^2+x+1\right)}$, The number of new coupling constants at each order would be this value minus $N-1$.}.
Similarly, we also have the freedom to redefine the auxiliary fields as $f_{N} \rightarrow f_{N} + a m^2 f_{N-1} + \ldots$ for some arbitrary constant $a$ (and likewise for $h_{N}$). Such field redefinitions can always be used to simplify or cancel terms appearing in $S_{2N}$ and $S_{2N+1}$. In the concrete examples presented in eqs.~\eqref{NMG} and \eqref{S3}, and the ones coming later, we have used such shifts to cancel the kinetic terms of the lower order action.

%
%  Not all couplings in \eqref{SNodd} are physical, for a given $N$ we have $2N+1$ auxiliary fields ($f_I$, $h_J$, $I=1,\cdots,N+1$ and $J=1,\cdots,N$)
% which can be rescaled such that $2N+1$ of terms are normalized to one. In \eqref{SNodd} we have already normalized arbitrarily
% $N+2$ kinetic terms by rescaling $N+2$ of fields\footnote{At each order $N$ the number of  interaction terms is given by the $x^{N}$ coefficient in the Taylor expansion of $\frac{x \left(x^3-x^2-x-1\right)}{(x-1)^3 \left(x^2+x+1\right)}$, The number of new coupling constants at each order would be this value minus $N-1$.}.
%

%In the parity-odd case, the CSG action has an additional (conformal) gauge symmetry, which is broken by the extra auxiliary fields. The resulting theory, with three auxiliary fields, propagates three modes: two massive helicity $\pm 2$ modes and a partially massless mode, as we will see in section \eqref{linearth}.

In this work we will only analyze extensions which preserve parity. It  is straightforward to extend the analysis to parity-violating models, such as TMG, by taking the sum of an even and odd parity theory. We will explicitly construct the even and odd parity extensions up to eight derivatives  in the metric formalism. In section \ref{linearth}, we will perform  the linearized analysis  and confirm that generically each set of auxiliary fields will add 2 massive spin-2 degrees of freedom. Before closing this section, we will comment on the absence of scalar ghosts and the growth of local degrees of freedom by adding each $(f,h)$-pair of auxiliary fields.

\subsection{Absence of scalar ghosts}\label{counting}

% In this section we first show how adding a pair of auxiliary fields $(f, h)$ increases the degree of freedom of the full non-linear theory by 2.
The advantage of the first order formulation over the metric form is that it is relatively easy to count the number of local degrees of freedom (dof) and identify the second class constraints which remove the Boulware-Deser scalar ghost. They arise from the symmetry of the auxiliary fields $h_{\mu\nu}$ and $f_{\mu\nu}$,
\begin{equation}\label{fhantisym}
f_{[\mu\nu]} = 0 \,, \qquad\qquad h_{[\mu\nu]} = 0\,.
\end{equation}
These constraints can be derived directly from the equations of motion \eqref{ENMGeom} by acting on them with an exterior derivative and using that $d^2 =0$. By invertibility of the dreibein, the first equation in \eqref{ENMGeom} simplifies to $f_{1\,a} e^a = 0$ and the second one gives $h_{1\,a} e^a = 0$, whose spatial projections are secondary constraints in a Hamiltonian formulation of the theory \cite{Hohm:2012vh,Bergshoeff:2014bia}.
The counting of degrees of freedom was shown in \cite{Hohm:2012vh} for NMG but it can be generalized to all CS-like theories considered in the preceding section.
The absence of additional scalar degrees of freedom then follows from a counting similar to the NMG case,
provided that the secondary constraints \eqref{fhantisym} are second class and do not lead to further tertiary constraints\,\footnote{
In this counting we assume that adding these new auxiliary fields does not change the number of gauge symmetries.
This is actually what happens  for conformal gravity; the presence of additional symmetries  cancels the degree of freedom introduced by $f^a$, see \cite{Afshar:2011qw}.}.
After a space-time decomposition of the fields, the time components $f^a_0$ and $h^a_0$, become Lagrange multipliers for a set of six primary constraints
and the spatial components of the fields, $f^a_i$ and $h^a_i$, add to the canonical variables of the theory. Along with the additional secondary constraints $f^a_{[ij]}=0$ and $h^a_{[ij]}=0$,
each pair of auxiliary fields will add
\begin{equation}\label{ENMGcounting}
\frac{1}{2}\left( 12 - 6 - 2\right)  = 2\
\end{equation}
degrees of freedom to the theory. These two degrees of freedom correspond to the two helicity states of a massive spin-2 mode in three dimensions. This counting works for all vector valued one-form pairs ($f_I$, $h_J$).
Hence, any action which gives the equations of motion with the general structure given in eq.~\eqref{ENMGeom} is guaranteed to produce a higher-derivative extension of gravity in three dimensions, free of scalar ghosts.

\section{Extended new massive gravity}\label{ExtendedNMG}
In this section we construct the extensions in the parity-even sector given in eq. \eqref{SNeven} up to eight derivatives. A similar analysis for the odd case is done in appendix \ref{ExtendedCSG} up to seven derivatives. Below we will first introduce the model. Next, in section
\ref{linearth} we will perform a linearized analysis of the model while in section \ref{crit}
we will investigate critical points and critical lines in the parameter space.
%  and do the linearized analysis in for the extended new massive gravity which is the six order derivative theory in subsection \cite{Linearization}

The  NMG action $S_2$ which is fourth order in derivatives was already given in eq.~\eqref{NMG}. The next step is to construct the
six derivative  action $S_4$. Its Lagrangian three-form\,\footnote{We define the Lagrangian three form $L$ and the Lagrangian density $\cL$ as $S=\frac{1}{\kappa^2}\int L=\frac{1}{\kappa^2}\int e\, d^3x\cL$, where $e$ denotes the determinant of the dreibein.} can be derived using the recursive action \eqref{SNeven}. We find the following result:
\begin{equation}\label{LENMG}
\begin{split}
 L_4 = - e_a&\left(\sigma  R^a - \frac{\Lambda_0}{6} \epsilon_{abc} e^b e^c \right) + \frac{1}{2m^2} \epsilon_{abc} e^af_1{}^b f_1{}^c  - \frac{1}{m^4}\Big[e_a \cD h_2{}^a    \\
 & + \frac{a}{6} \epsilon_{abc} f_1{}^a f_1{}^b f_1{}^c + f_{2\,a} \left( R^a + \epsilon^{abc}e_b f_{1\,c} \right) + b\,  h_{1\,a}\left(\cD f_1{}^a + \tfrac{1}{2} \epsilon^{abc} e_b h_{1\,c} \right)\Big]  \,,
\end{split}
\end{equation}
where we have introduced a sign parameter $\sigma = \pm 1$ and two arbitrary dimensionless parameters $a,b $. The dimensionful parameters $\Lambda_0$  and $m^2$ were already introduced  in eqs.~\eqref{EHG} and \eqref{NMG}.

The equations of motions for this Lagrangian, obtained by varying with respect to $h_2{}^a$, $f_2{}^a$, $h_1{}^a$, $f_1{}^a$, $\omega^a$ and $e^a$, respectively, are given by
\begin{equation}\label{ENMGeom2}
\begin{split}
& \cD e^a = 0\,,\\
& R^a + \epsilon^{abc} e_b f_{1\,c} = 0\,,\\
& \cD f_1{}^a + \epsilon^{abc} e_b h_{1\,c} = 0\,, \\
& b\, \cD h_1{}^a + \tfrac{1}{2} \epsilon^{abc} \left( a \,f_{1\,b} f_{1\,c}+2 e_b f_{2\,c} -  2m^2 e_b f_{1\,c}  \right) = 0\,,  \\
& \cD f_2{}^a + \epsilon^{abc} \left( b\,f_{1\,b} h_{1\,c} + e_b h_{2\,c} \right) =0\,,  \\
& \cD h_2{}^a + \tfrac{1}{2} \epsilon^{abc} \left( 2f_{2\,b} f_{1\,c} + b\, h_{1\,b} h_{1\,c} - m^2 f_{1\,b} f_{1\,c} - \Lambda_0 m^4 e_b e_c - 2m^4 \sigma \, e_b f_{1\,c} \right) = 0 \,.
\end{split}
\end{equation}
The first equation imposes the zero torsion constraint and allows the spin-connection to be solved for  in terms of the dreibein. Moving down the line, we find for the fields $f_{1\, \mu\nu}$ and $h_{1\,\mu\nu}$ the
following expressions:
\begin{equation}
f_{1\,\mu\nu}= -S_{\mu\nu}(e) \qquad \text{and} \qquad
h_{1\, \mu\nu} =  C_{\mu\nu}(e)\,. \label{f1h1}
\end{equation}
These in turn fix the expressions for  $f_{2\,\mu\nu}$ and $h_{2\,\mu\nu}$ as follows:
\begin{align}
f_{2\,\mu\nu} &=  - b \,D_{\mu\nu} + a \left( P_{\mu\nu}-\tfrac{1}{4}P  g_{\mu\nu}  \right)  - m^2 S_{\mu\nu}\,,\\
h_{2\, \mu\nu} &=  - E_{\mu\nu} - 2 b\left(   Q_{\mu\nu} - \tfrac{1}{4}  Q g_{\mu\nu}\right) + b\,SC_{\mu\nu}  \,,
\end{align}
where $D,E,P$ and $Q$ are defined by
\begin{align}\label{DPEQten}
D_{\mu\nu}&\equiv e^{-1} \epsilon_{(\mu|}{}^{\alpha\beta} \nabla_{\alpha} C_{\beta |\nu)}\,,\;\qquad
P_{\mu\nu}\equiv G_{\mu}{}^{\rho} S_{\nu \rho}\,,\\
E_{\mu\nu}&\equiv e^{-1} \epsilon_{(\mu|}{}^{\alpha\beta} \nabla_{\alpha} f_{2\,\beta |\nu)}
\,,\qquad Q_{\mu\nu}\equiv C_{(\mu}{}^{\rho} S_{\nu) \rho}\,.
\end{align}
Substituting these expressions back into  the action leads to the following  `extended' NMG (ENMG) Lagrangian density
\begin{equation} \label{NMG2}
 \mathcal{L}_{\text{\tiny{ENMG}}}\equiv\mathcal{L}_4= \frac{1}{2}\left\{\sigma R-2\Lambda_0 + \frac{1}{m^2} P + \frac{1}{m^4}\left( 2 a \det(S) - b \,C^{\mu\nu}C_{\mu\nu}\right)\right\}\,.
\end{equation}
% where $a$ and $b$ are arbitrary parameters and
% 
At order $1/m^2$ we have the NMG combination of $R^2$ terms,
\bea
P =R_{\mu\nu}R^{\mu\nu}-\frac{3}{8}R^2,
\eea
while at order $1/m^4$ we find the following two combinations of six-derivative  terms:
\begin{equation}
\begin{split}\label{Cubic}
- 6 \det(S)& = 2R_{\mu}^{\nu}R_{\nu}^{\rho}R_{\rho}^{\mu}-\frac{9}{4}RR_{\mu\nu}R^{\mu\nu}+\frac{17}{32}R^3\,, \\
C^{\mu\nu}C_{\mu\nu}& = R_{\mu\nu}\square R^{\mu\nu} - \frac{3}{8}R\square R - 3R_{\mu}^{\nu}R_{\nu}^{\rho}R_{\rho}^{\mu} + \frac{5}{2}RR_{\mu\nu}R^{\mu\nu} - \frac{1}{2}R^3\,.
\end{split}
\end{equation}
The last identity is up to total derivatives. This theory is free of scalar ghosts and has four local degrees of freedom by construction, as was verified in section \ref{counting}.

We can systematically continue this program and find more and more higher order  terms. As
an example we present the eight-derivative  theory:
\begin{align}\label{EENMG}
 \mathcal{L}_6=
\mathcal{L}_{\text{\tiny{ENMG}}}+\frac{1}{m^6}\bigg\{&\kappa_1\left(P_{\mu\nu}P^{\mu\nu}-\tfrac{3}{8}P^2\right)+\kappa_2\left( S_{\mu}^ \rho C_{\rho\nu} C^{\mu\nu}-\tfrac{1}{2}SC_{\mu\nu}C^{\mu\nu}\right)\nn\\
&+\kappa_3\left(C_{\mu\nu}\square C^{\mu\nu}+3S_{\mu}^ \rho C_{\rho\nu}  C^{\mu\nu} +SC_{\mu\nu}C^{\mu\nu}\right)\bigg\}\,,
\end{align}
with parameters $\kappa_1,\kappa_2$ and $\kappa_3$. One can simplify the $\kappa_1$-term using the Schouten identity $\slashed{ S}_{\mu\nu}^4=\frac{1}{2}(\slashed{S}_{\mu\nu}^2)^2$, where
$\slashed{S}$ is the traceless Schouten tensor, $\slashed{S}_{\mu\nu}=S_{\mu\nu}-\frac{1}{3}Sg_{\mu\nu}$, as follows:
\bea
-\frac{\kappa_1}{12}\left(16\,SS_\mu^\rho S_\rho^\sigma S_\sigma^\mu -3(S_{\mu\nu}S^{\mu\nu})^2-18S_{\mu\nu}S^{\mu\nu}S^2+5S^4\right)\,.
\eea

The $a$- and $\kappa_1$-terms above are precisely the combination of $R^3$ and $R^4$ terms found by Sinha in \cite{Sinha:2010ai} by demanding the presence of a holographic $c$-theorem in higher-derivative extensions of new massive gravity, see also \cite{Paulos:2010ke}. The $b$-,  $\kappa_2$- and $\kappa_3$-terms were not considered in their considerations regarding the holographic $c$-theorem. We will comment on this in section \ref{sec:ctheorem}.

% \newpage

\subsection{Linearization}\label{linearth}
In this section we study the extended NMG theory  by linearizing the model
around a maximally symmetric vacuum parametrized by a background dreibein $\bar{e}$,  spin-connection $\bar{\omega}$ and cosmological constant $\Lambda$ that satisfy
\begin{equation}\label{vacuum}
\bar{R}^a \equiv\bar{\mathcal D}\bar \omega^a= \frac{\Lambda}{2} \epsilon^{a}{}_{bc} \bar{e}^b \bar{e}^c\,,\qquad \qquad \bar{\cD}\bar{e}^a = 0\,.
\end{equation}
All barred quantities refer to the background. The background values for the auxiliary fields can be determined by their background equations of motion. Since the parity-even and parity-odd models have the same field equations,
these fields have the same background values in both models. The $f$ fields all have background values proportional to the background dreibein. The $h$ fields, which are constructed from the Cotton tensor, vanish on this background.
We parametrize the vector-valued fluctuations of the one-forms around the vacuum as
\begin{align}\label{fluctuations}
    e &= \bar{e} + \kappa \,k_0\,, & \omega &= \bar{\omega} + \kappa\, v_0\,, \nonumber \\
    f_1{} &= -\tfrac{\Lambda}{2}\,\bar{e}  + \kappa \,k_1{} \,, &      h_1{} &= \kappa\, v_1{} \,, \\ \nonumber
    f_2{} &=-\tfrac{\Lambda}{2} \left( m^2 + \tfrac{a \Lambda}{4} \right)  \bar{e}  + \kappa \,k_2{} \,, &      h_2 &= \kappa\, v_2{} \, ,\\
    &\vdots & \vdots &\,\nonumber
%     \nonumber   &\cdots    &    &\cdots
\end{align}
where we used $\kappa$ as a small expansion parameter.
%In the parity-odd case $m$ and $a$ are replaced by $\mu$ and $\alpha$.
We next  substitute eq.~\eqref{fluctuations} into the recursive action \eqref{SNeven}  and keep the quadratic terms which are $\kappa$-independent.

 We first focus on the quadratic Lagrangians $L_0^{(2)}$ and $L_1^{(2)}$ corresponding to the Einstein gravity action  \eqref{EHG} and the conformal gravity action  \eqref{CSG} which have no propagating degrees of freedom. Plugging the expressions \eqref{fluctuations} into the actions \eqref{EHG} and \eqref{CSG} we find
\begin{align}\label{LEHG}
L_0^{(2)} = &- \left\{ k_{a} \bar{\cD} v^a + \frac12 \epsilon_{abc}\, \bar{e}^a \left( v^bv^c - \Lambda k^b k^c \right)\right\}  \,,\quad\Lambda=\Lambda_0\,,\\ 
L_1^{(2)} = &\,\mu^{-1} \left\{    k_{1\,a}\left(\bar{\cD} k^a + \epsilon_{abc}\,\bar e^b v^c \right)+\frac12 v_{a} \left(\bar{\cD} v^a - \Lambda \epsilon_{abc}\,\bar e^b k^c\right) \right\}  \,,\label{LCSG}
\end{align}
where we have defined $k\equiv k_0$ and $v\equiv v_0$.
By using the field equations for $v^a$ and $k_1^a$ it is possible to eliminate them in $L_0^{(2)}$ and $L_1^{(2)}$. We thus obtain the quadratic Lagrangian density in the following  more conventional second order form:
\begin{align}
\cL_0^{(2)} =  -  k^{\mu\nu} \cG_{\mu\nu}(k) \,\qquad\text{and}\qquad\cL_1^{(2)} = -  \frac{1}{\mu}\epsilon^{\mu\alpha}{}_{\rho} \bar{\nabla}_{\alpha} k^{\rho\nu} \cG_{\mu\nu}(k) \,.
\end{align}
The corresponding linear field equations are given by
\begin{equation}
\cG_{\mu\nu}(k) = 0\,\qquad\text{and}\qquad( \cD^0 \cG(k))_{\mu\nu} = 0\,.
\end{equation}
Here $\cG_{\mu\nu}(k)$ is the linearized Einstein tensor invariant under linear diffeomorphisms. In the transverse traceless gauge we have,
\bea\label{linG}
\cG_{\mu\nu}(k)= %\frac{1}{2}(\cD^L \cD^R k)_{\mu\nu}=
-\frac{1}{2}\left(\bar{\square}-2\Lambda\right)k_{\mu\nu}\,.
\eea
% where
% \bea\label{DLDRopt}
% (\cD^{L/R})_{\mu}^{\rho} = \delta_{\mu}^{\rho} \pm  \ell \epsilon_{\mu}{}^{\alpha\rho} \bar{\nabla}_{\alpha}\,.
%  \eea
Clearly, $\mathcal{L}_0^{(2)}$ is the linearized Einstein-Hilbert term without any propagating degrees of freedom. On the other hand, in $\cL_1^{(2)}$  there is an additional partially massless mode satisfying 
\begin{equation}
(\cD^0k)_{\mu\nu} = \epsilon_{\mu}{}^{\alpha\rho} \bar{\nabla}_{\alpha}k_{\rho\nu}=0 \,.\label{DOopt}
\end{equation}
This equation is invariant  under an additional linearized Weyl transformation \cite{Afshar:2011qw}.
Below, we discuss the linearized theories of the parity-even sector up to $S_4$ and confirm that the addition of two auxiliary fields adds a massive spin-2 mode. For a similar analysis of the parity-odd sector  we refer  to appendix \ref{ExtendedCSG}.

% \subsection{Parity Even Models}
Moving to NMG, we consider the four derivative action $S_2$, see eq.~\eqref{NMG}, where we have included the auxiliary form fields $f_1$ and $h_1$. The background field equation enforces the following quadratic relation for $\Lambda$:
\begin{equation}
\Lambda_0 = \Lambda\left(\sigma  + \frac{\Lambda}{4m^2}\right)\,.
\end{equation}
Making the field redefinition, $k_1\rightarrow k_1-\frac{\Lambda}{2}k$, the quadratic Lagrangian three-form reads,
\begin{equation}
\begin{split}
L_2^{(2)} =  \left(\sigma - \frac{\Lambda}{2m^2} \right) L_0^{(2)} - \frac{1}{m^2}  \bigg\{
& k_{1\,a} \left( \bar{\cD} v^a - \Lambda \epsilon^a{}_{bc} \bar{e}^b k^c + \tfrac12  \epsilon^a{}_{bc} \bar{e}^b k_1^c \right)
 \\ & +  v_{1\,a} \left(  \bar{\cD} k^a + \epsilon^a{}_{bc} \bar{e}^b v^c \right) \bigg\} \,,
 \end{split}
\end{equation}
where $L_0^{(2)}$ is given in eq.~\eqref{LEHG}.
Eliminating $v^a$ and $v_1^a$ using their equations of motion, the Lagrangian reduces to the following density:
\begin{align}\label{LNMG}
\cL_2^{(2)} = &\, -\sigma_2  k^{\mu\nu} \cG_{\mu\nu}(k)- \frac{2}{m^2}  k_1^{\mu\nu} \cG_{\mu\nu}(k) - \frac{1}{2m^2}( k_{1\,\mu\nu} k_1{}^{\mu\nu} - k_1^2)\,,\nn\\
= & \,\sigma_2 \,\cL_0^{(2)} - \frac{1}{m^4\sigma_2} \cL_{\text{\tiny{FP}}}(k_1,\cM) \,,\qquad\sigma_2=\sigma - \frac{\Lambda}{2m^2} =-\frac{\cM^2}{m^2}\,.
\end{align}
The  Fierz-Pauli Lagrangian $\cL_{\text{\tiny{FP}}}$ is given by
 \bea
\cL_{\text{\tiny{FP}}}(\tilde{k},\cM)= -\tilde k^{\mu\nu}\mathcal{G}_{\mu\nu}(\tilde k)-\frac{1}{2}\cM^2(\tilde k^{\mu\nu}\tilde k_{\mu\nu}-\tilde k^2)  \,,
\eea
for a generic massive mode $\tilde k$. Assuming that $\sigma_2\ne0$,  the Lagrangian density $\cL_2^{(2)}$
has been diagonalized by a proper shift in the $k$ field;
\begin{equation}
k_{\mu\nu} \rightarrow k_{\mu\nu}+ \frac{1}{m^2 \sigma_2}\, k_{1\,\mu\nu}\,.
\end{equation}
To avoid tachyons the mass of the massive mode should  be bounded as $\mathcal M^2\geq0$ (see for instance \cite{Bergshoeff:2010iy,Bergshoeff:2014pca}). In order to avoid ghosts both kinetic terms in \eqref{LNMG} should come with the same sign. This already shows that depending on the sign of $\sigma_2$, either the massless mode or the massive mode in \eqref{LNMG}  is ghost-like. The massless mode should not be ghost-like as it determines the charges for the gravitational theory through the Brown-York stress tensor. We therefore take $\sigma_2>0$. However,  restricting to $\sigma_2>0$ results in a massive spin-2 ghost. As we will see later this is a general problem in higher-derivative theories, denoted as the bulk-boundary unitarity clash.

We  now go one step further and consider the quadratic part of the action $S_4$. The cosmological parameter $\Lambda_0 $ is now related to the physical cosmological constant $\Lambda$ by the cubic equation
\begin{equation}\label{ENMGbkgd}
 \Lambda_0 = \Lambda\left(\sigma  + \frac{\Lambda}{4 m^2} + \frac{a \Lambda^2}{8 m^4}\right)\,.
\end{equation}
By performing  the field redefinition
\begin{align}
k_2\rightarrow k_2-\tfrac{\Lambda}{2}(m^2+\tfrac{a\Lambda}{4})k\,,\qquad
k_1\rightarrow k_1-\tfrac{\Lambda}{2}k\,,\qquad
v_2\rightarrow v_2+\tfrac{b\Lambda}{2}v_1\,,
\end{align}
the quadratic Lagrangian three-form part  of $S_4$ can be brought into the following form:
\begin{equation}
\begin{split}
L_4^{(2)} = & \; \left(\sigma - \frac{\Lambda}{2m^2} - \frac{a\Lambda^2}{8m^4} \right)  L_0^{(2)} + \frac{1}{2m^2}\left(1 + \frac{a \Lambda}{2m^2} \right) \epsilon_{abc} \bar{e}^a k_1{}^b k_1^c \\
-& \frac{1}{m^4}\bigg\{  k_{2\,a} \left( \bar{\cD} v^a - \Lambda \epsilon^a{}_{bc} \bar{e}^b k^c +  \epsilon^a{}_{bc} \bar{e}^b k_1^c \right)
  +  v_{2\,a} \left(  \bar{\cD} k^a + \epsilon^a{}_{bc} \bar{e}^b v^c \right)
 \\ &\qquad- b \,v_{1\,a} \left(  \bar{\cD} k_1^a +  \tfrac12 \epsilon^a{}_{bc} \bar{e}^b v_1{}^c \right)   \bigg\} \,.
 \end{split}
\end{equation}
Upon eliminating the auxiliary fields $v{}^a, v_1{}^a$ and $v_2{}^a$ by using their equations of motion, the quadratic Lagrangian density may be written as
\begin{align} \label{linShk}
\cL_4^{(2)} = &\, -\sigma_4 \, k^{\mu\nu} \cG_{\mu\nu}(k)  - \frac{2}{m^4} k^{\mu\nu}_2 \cG_{\mu\nu}(k) -  \frac{b}{m^4} k_1^{\mu\nu} \cG_{\mu\nu}(k_1) \nonumber \\
& + \frac{1}{2m^2}\,\Theta \left(k_1^{\mu\nu} k_{1\mu\nu} - k_1^2\right) - \frac{1}{m^4} \left( k_1^{\mu\nu}k_{2\mu\nu} - k_1 k_2 \right)  \,,
\end{align}
where
\bea
\sigma_4=\sigma - \frac{\Lambda}{2m^2}- \frac{a\Lambda^2}{8m^4}\,\qquad\text{and}\qquad\Theta=1  + \frac{a \Lambda}{2m^2} - \frac{b\Lambda}{ m^2} \,.
\eea
For general values of the parameters this quadratic Lagrangian leads to a sixth order differential equation for $k_{\mu\nu}$.
The matrix for the kinetic terms and the mass terms in the basis defined by $|k\rangle$, $m^2|k_1\rangle$ and $m^4|k_2\rangle$ can be written as,
\bea\label{2matrices}
\mathcal K=-\left( \begin{array}{ccc}
\sigma_4  & 0 & 1 \\
0 & b & 0 \\
1 & 0 & 0 \end{array} \right)\,\qquad\text{and}\qquad
\mathcal M^2=m^2\left( \begin{array}{ccc}
0  & 0 & 0 \\
0 & -\Theta & 1 \\
0 & 1 & 0 \end{array}\right)\,.
\eea
Assuming that $\sigma_4 \ne 0$ and $b\ne 0$  these two matrices can be diagonalized simultaneously by redefining the fields
\begin{align}
k & = k^0 - \frac{1}{\sigma_4} \left(k^+ - k^-\right)\,, \\
k_1 & = - \frac{m^4}{b }\left(\frac{1}{ \cM_-^2} k^+- \frac{1}{\cM_+^2}  k^-\right) \,, \\
k_2 & = m^4 \left( k^+ -  k^- \right)\,,
\end{align}
such that the quadratic Lagrangian becomes the sum of the linearized Einstein-Hilbert term and two Fierz-Pauli terms
\begin{align}
\cL_4^{(2)}&= \sigma_4 \cL_{0}^{(2)}(k^0) + \mathcal K_+ \cL_{\text{\tiny{FP}}}(k^+,\mathcal M_+) + \mathcal K_-\,\cL_{\text{\tiny{FP}}}(k^-,\mathcal M_-)\,.
\end{align}
Here $\sigma_4$ and $\mathcal K_{\pm}$ satisfy
\begin{equation}\label{KK}
\sigma_4\,\mathcal K_{+} \mathcal K_{-}= -\frac{\Theta^2-4 b\, \sigma_4}{b \, \sigma_4^2} = \mathcal K_{+}+ \mathcal K_{-}\,,
\end{equation}
while the Fierz-Pauli masses $\cM_{\pm}^2$ are given by
\bea\label{MM}
\cM_+^2 \cM_-^2=\frac{m^4}{b}\sigma_4  \,\qquad\text{and}\qquad \cM_+^2 - \cM_-^2= \frac{m^2}{b} \sqrt{\Theta^2 - 4 b \sigma_4}\,.
\eea
The numerator in \eqref{KK} should be positive, otherwise the masses in \eqref{MM} become imaginary.
 Like in the previous case, we see that it is not possible to achieve a positive sign for the kinetic terms and the masses simultaneously. From \eqref{KK} and \eqref{MM} we see that  there is always  either a negative mass squared (when $b\sigma_4 < 0$) or a wrong-sign kinetic term (when $b\sigma_4 > 0$) in the theory. %For $\sigma_4 < 0$ one ($b<0$) or both ($b>0$) kinetic terms have the wrong sign. 
 For all values where $\sigma_4 \neq 0$ or $b\neq0$, one of the massive modes is either tachyonic or a ghost. In the next section we will discuss what happens at special points in the parameter space.

\subsection{Critical lines and the tricritical point}\label{crit}
In the above analysis we disregarded the points in the parameter space that reduce the rank of the
two matrices given in eq.~\eqref{2matrices}. Below we discuss these special points separately.
\begin{enumerate}
\item[] {$\boldsymbol{b=0}$}: At this point  the rank of the matrix $\mathcal K$ is reduced by one. The action \eqref{LENMG} is now independent of the auxiliary field $h_1$, and $f_2$ is algebraically given in terms of $f_1$. This reduces the  number of local degrees of freedom from four to two, representing a single massive graviton. From eq.~\eqref{NMG2} we see that the term involving the Cotton tensor has disappeared and the action reduces to the `cubic extended'  NMG model described in \cite{Sinha:2010ai}. 

\item[] {$\boldsymbol{b=\Theta=0}$}: At this special point the linearized equations become second order in derivatives and the massive mode disappears from the linearized spectrum. Note that there are no ghosts left in the linearized theory \cite{Sinha:2010ai}. It is however not clear if this feature survives at the non-linear level. 

\item[] {$\boldsymbol{\sigma_4=0}$}:
At this critical line %$a = a_{\rm crit} = -4 \Lambda^{-1} m^2(1 - 2 \Lambda^{-1} m^2 \sigma)$
one of the FP masses vanishes, but the linearized equations remain sixth-order in derivatives. Consequently a new, logarithmic (log)-mode appears and together with the massless mode it forms a Jordan cell of rank two.
The Lagrangian \eqref{linShk} is not diagonalizable any more.

\item[] {$\boldsymbol{\sigma_4=\Theta=0}$}:
This is a `tricritical' point, % $b = b_{\rm crit} = -\Lambda^{-1} m^2(1-4\Lambda^{-1} m^2 \sigma)$
where both FP masses vanish and  the corresponding massless gravitons form a Jordan cell of rank three. The spectrum now contains one log-mode and a log$^2$-mode (see for instance \cite{Bergshoeff:2012ev,Grumiller:2010tj}).

\item[] {$\boldsymbol{\Theta^2 = 4 b \sigma_4}$}:
This is another critical line where the two non-zero FP masses degenerate and form a Jordan cell of rank two. At this point the spectrum contains one massive mode and a massive log-mode.
\end{enumerate}
Among the above critical points, $\sigma_4=0$ and $\Theta=0$ can only occur when $\Lambda\neq0$.  They are interesting from the AdS/CFT point of view. Especially the $b=\Theta=0$ point is interesting as the linearized analysis suggests perturbative unitarity. 
For a  more detailed
 treatment of the $\sigma_4=0$ case, see section \ref{LCFT} and also the reference \cite{Bergshoeff:2012ev} where most of the computations performed in the context of the AdS/LCFT correspondence carry over to this model.%\footnote{Note that in this case the parameter space is larger than the models discussed in \cite{Bergshoeff:2012ev}. In fact, after tuning $a$ and $b$ to their critical values, $m^2$ remains a free parameter and hence the tricritical point is a continuous collection of points in parameter space. We expect this to be reflected in the new anomaly at the tricritical point by the appearance of $m^2$ in its value, in contrast to the fixed combination of $\ell/G$ that was found in \cite{Bergshoeff:2012ev}.}

\section{Anti-de~Sitter holography}\label{Holography}
All the extended massive gravity models we constructed so far admit an AdS vacuum. Hence, it is possible to study their holographic dual by imposing suitable asymptotically AdS boundary conditions. The asymptotic symmetry algebra emerges as the algebra of conserved global charges related to gauge transformations which preserve the AdS boundary conditions. The procedure for finding the asymptotic symmetry algebra is similar to that of pure CS gauge theories on manifolds with a boundary \cite{Banados:1994tn}. The main difference is that for CS-like theories not all constraint functions generate gauge symmetries. The first step is to identify the first class constraints and their corresponding boundary terms for the CS-like theories we have discussed in the preceding section. This is done in full detail in appendix \ref{can}. Here we will briefly summarize the main results derived there and continue with a discussion on the suitable boundary conditions and the asymptotic symmetry 
transformations which preserve them. After deriving the central charge in the  asymptotic symmetry algebra of extended NMG, we discuss the appearance of Jordan cells at special points in its parameter space where the central charge vanishes and compute the new logarithmic anomalies. We conclude this section by showing how the models constructed in this paper are consistent with a holographic $c$-theorem.

\subsection{Gauge symmetries in CS-like theories}%Algebra of first class constraints

In appendix \ref{can} we have identified, using standard techniques, the first class constraints $\phi_{\text{LL}}$ and $\phi_{\text{diff}}$ of a general CS-like model that generate local Lorentz transformations  and diffeomorphisms, see eqs.~\eqref{phiLL} and \eqref{phidiff}, respectively. On the AdS background, these can be written in terms of a set of mutually commuting $SL(2,\mathbb{R})$ generators $J_{\pm}$ with Poisson bracket algebra
\begin{equation}\label{Lbrack}
\{ J_{\pm} [\xi], J_{\pm} [\eta] \} = J_{\pm}[\xi \times \eta] +\,\text{\small{B.T.}}\,, \qquad \{J_+[\xi],J_{-}[\eta]\} = 0\,.
\end{equation}
In general, the presence of a boundary introduces non-trivial boundary terms in the definition of the $J_{\pm}$ and in the Poisson bracket algebra \eqref{Lbrack}. In appendix \ref{can} we show that for the CS-like theories we consider here, the improved generators $\mathcal{J}_\pm$ defined by
\begin{equation}
\mathcal{J}_\pm[\xi^{\pm}] = J_{\pm}[\xi^{\pm}] + Q_{\pm}[\xi^{\pm}]\,,
\end{equation}
are differentiable provided that the variation of the boundary term $Q_{\pm}$ takes the form
\begin{equation}\label{delQ}
\delta Q_{\pm}[\xi^{\pm}] =  \frac{\hat k}{2\pi} \int_{\partial \Sigma} dx^i \; \xi_a^{\pm} \left( \delta \omega_i{}^a \pm \frac{1}{\ell} \delta e_i{}^a \right)\,.
\end{equation}
Here $\hat{k}$ is an effective coupling which depends on the specific theory under consideration. The Poisson brackets of the improved generators then pick up a boundary term which can be derived from the general formula \eqref{gen_poissonbr} in appendix \ref{can}:
\begin{equation}\label{bdyterms}
\{\mathcal{J}_{\pm}[\xi^{\pm}], \mathcal{J}_{\pm}[\eta^{\pm}]\} = \cdots + \frac{\hat k}{4\pi } \int_{\partial \Sigma} dx^i \; \xi^{\pm\,a} \left[ \partial_i \eta^{\pm}_a + \epsilon_{abc} \left(\omega_i{}^b \pm \frac{1}{\ell} e_i{}^b \right) \eta^{\pm\,c} \right]\,.
\end{equation}
Here the dots denote bulk terms. After adopting suitable boundary conditions, the charges become integrable and the above boundary term will provide a term needed to improve the bulk part in \eqref{Lbrack} and a central extension. We can now discuss the AdS boundary conditions.

\subsection{AdS boundary conditions and the central charge}\label{sec:cc}
To give the boundary conditions, it is convenient to  represent the spin-connection and the dreibein of the AdS background with radius $\ell$ in the following combinations:
\begin{align}\label{ADS}
\bar{\omega}+\frac{\bar{e}}{\ell}=&\,b^{-1}\Big(L_1 +\tfrac{1}{4}L_{-1}\Big)b\,dx^++b^{-1}\partial_\rho b\,d\rho\,,\nn\\
\bar{\omega}-\frac{\bar{e}}{\ell}=&-b\Big(L_{-1}+\tfrac{1}{4}L_{1}\Big)b^{-1}dx^-+b\,\partial_\rho b^{-1}d\rho\,,
\end{align}
where $x^\pm=\frac{t}{\ell}\pm\varphi$, $b=e^{\rho L_0}$ and,
\bea\label{Lorentz}
L_1=J_0+J_1\,,\qquad L_{-1}=J_0-J_1\,\qquad\text{and}\quad L_0=J_2\,.
\eea
The AdS boundary conditions are  presented in terms of some free state dependent normalizable contributions to this background. These contributions behave as the vacuum expectation value (VEV) for the boundary operators
which are sourced by non-normalizable modes. As we discussed before, each pair ($f_I, h_I$) introduces two new degrees of freedom representing a massive spin-2 normalizable mode with mass $\mathcal M^2$ which satisfy the following
equation,
\bea
\tilde{\mathcal{D}}^M\mathcal{D}^Mk^I_{\mu\nu}=0\,,
\eea
 with
\bea\label{DMopt}
\big({\mathcal D}^{M}\big)_\mu^\nu = \delta_\mu^\nu + \frac{1}{M}\,\epsilon_\mu{}^{\tau\nu}\bar\nabla_\tau\,,\qquad\text\qquad \big(\tilde{\mathcal D}^{M}\big)_\mu^\nu = \delta_\mu^\nu - \frac{1}{M}\,\epsilon_\mu{}^{\tau\nu}\bar\nabla_\tau
\eea
 and $M^2\ell^2=\mathcal M^2\ell^2 +1$. The non-normalizable partner of this mode plays the role of a source which couples to a new operator $\mathcal O_I$
in the dual conformal field theory with conformal weights $(h,\tilde h)$. These  weights  are related to the mass and angular momentum of the bulk mode via the relations \cite{Li:2008dq}
\bea
\Delta=h+\tilde h=1+|M\ell|\;\qquad\text{and}\qquad \;s=h-\tilde h=\pm2\,,
\eea
  where $\pm$ distinguishes between the right and left sectors. The requirement of unitarity bounds the scaling dimension $\Delta$ as $\Delta\ge |s|$ (when   $1\le\Delta<|s|$   one of the conformal weights $(h,\tilde h)$ is negative). When the masses of $N$  modes degenerate, the conformal weights of their corresponding operators also degenerate and they form a Jordan cell of rank-$N$  in a logarithmic conformal field theory.
  
In this work we are interested in the asymptotic symmetry algebra generated by the gauge symmetries of the bulk theory. For this reason, we will not consider the sources for the massive modes. The resulting AdS boundary conditions are called Brown-Henneaux boundary conditions. These boundary conditions are sufficient for finding the Poisson brackets between the gauge generators \cite{Carlip:2008qh}. In the first order formalism, inspired by the Chern-Simons formulation of 3D gravity, the Brown-Henneaux  boundary conditions on the dreibein and the spin-connection are given by\cite{Coussaert:1995zp}%,Banados:1994tn
\begin{align}\label{bcone}
\omega+\frac{e}{\ell}=&\,b^{-1}\left\{\Big(L_1+\mathcal L(x^+)L_{-1}\Big)dx^++\text{d}\right\}b\,,\nn\\
\omega-\frac{e}{\ell}=&-b\left\{\left(L_{-1}+\tilde{\mathcal L}(x^-)L_{1}\right)dx^--\text{d}\right\}b^{-1}\,.
\end{align}
The state dependent functions $\mathcal L(x^+)$ and $\tilde{\mathcal L}(x^-)$ are the vacuum expectation value (VEV) of the boundary energy-momentum operator.

% \subsection{Central Charge}\label{sec:cc}

%To calculate the central charge, we need the boundary charges $Q_{\pm}$ of the generators $L_{\pm}[\xi]$, which are given by the general formula \eqref{gen_varbc} in appendix \ref{can}. The variation of these
% boundary charges is given by
%\begin{equation}
%\begin{split}
%\delta Q_{\pm} = & - \int_{\partial \Sigma} dx^i \; \left( \xi^r_a g_{rs} + a_{\pm} \xi^e_a g_{\omega s}\right) \delta a_{i}^{s\,a} \,,\\
%=& - \int_{\partial \Sigma} dx^i \; \left(  g_{es} + \bar{f}_1 g_{f_1s} + \bar{f}_2 g_{f_2s} + \ldots + a_{\pm} g_{\omega s}\right) \xi^e_a\delta a_{i}^{s\,a}\,,
%\end{split}
%\end{equation}
%where in the first line the sum over $r$ does not include $\omega$ and where we have used eq.~\eqref{gpbkgd} in the last line. In general, after plugging in the explicit flavor space metric and AdS background values of the fields, the result may be written as
%\begin{equation}
%\delta Q_{\pm}[\xi^{\pm}] =  \frac{\hat k}{2\pi} \int_{\partial \Sigma} dx^i \; \xi_a^{\pm} \left( \delta \omega_i{}^a \pm \frac{1}{\ell} \delta e_i{}^a \right)\,.
%\end{equation}
%where $\hat k$ is an effective coupling determined by the elements of $g_{rs}$ and the $\bar{f}_I$'s. We have also distinguished the gauge parameters for the left and right moving sectors explicitly.

In order to integrate the expression \eqref{delQ} to the boundary charges, we impose the Brown-Henneaux boundary conditions \eqref{bcone} on the dreibein and the spin-connection.
The  gauge transformations preserving these boundary conditions are then given by:
\begin{equation}\label{gaugegen}
\xi^+ = b^{-1}\epsilon(x^+)b  \qquad\text{and}\qquad \xi^- = b\,\tilde\epsilon(x^-)b^{-1}\,,
\end{equation}
where\,\footnote{Since we are dealing with parity preserving models, we concentrate only on the left sector. The right sector will be determined
via a parity transformation.}
\bea
\epsilon^{-1}=\tfrac{1}{2}\epsilon''+\epsilon\mathcal L\,,\quad\epsilon^{0}=-\epsilon'\,\quad\text{with}\quad \epsilon^1\equiv\epsilon\,.
\eea
The variation of the state-dependent function $\mathcal{L}$ in eq.~\eqref{bcone} with respect to $\epsilon$, the parameter of the symmetry transformation, is given by,
\bea
\delta_\epsilon\mathcal{L}(x^+)=\epsilon(x^+)\mathcal{L}'(x^+)+2\epsilon'(x^+)\mathcal{L}(x^+)+
\tfrac{1}{2}\epsilon'''(x^+)\,.
\eea
% and the same for the bared sector.
This leads to the following expression for the conserved charge $Q = Q_{+} + Q_{-}$ at the boundary:
\begin{equation}
Q = \frac{\hat k}{2\pi}\int d\varphi \left[\epsilon(x^+)\mathcal{L}(x^+)-\tilde{\epsilon}(x^-)\tilde{\mathcal{L}}(x^-)\right]\,.
\end{equation}
We can now compute the Poisson brackets \eqref{Lbrack} with the boundary term \eqref{bdyterms} after suitably identifying $\xi$ and $\eta$ using eq.~\eqref{gaugegen} and defining the generators as,
\bea
L_n=\mathcal{J}_+[\epsilon=e^{inx^+}]\qquad\text{and}\qquad \tilde L_n=\mathcal{J}_-[\tilde \epsilon=e^{inx^-}]\,.
\eea
As expected we find two copies of Virasoro algebra,
\begin{align}
 i\{L_m,L_n\}&=(m-n)L_{m+n}+\frac{c}{12}m(m^2-1)\delta_{m+n,0}\,,\\
 i\{\tilde L_m,\tilde L_n\}&=(m-n)\tilde L_{m+n}+\frac{\tilde c}{12}m(m^2-1)\delta_{m+n,0}\,,
\end{align}
where $c=\tilde c=6\hat k$ for parity-even models and $c = - \tilde{c} = 6 \hat k$ for the parity-odd theories.
 %After explicitly computing the effective couplings $\hat{k}$ for ECSG (see appendix \ref{ExtendedCSG}) we find the central charges
%\begin{equation}
%c_{\text{\tiny ECSG}}  = - \tilde{c}_{\text{\tiny ECSG}} =   \frac{3}{2\mu G} \,.
%\end{equation}
%This result is independent of the new coupling constant $\alpha$.
In appendix \ref{can} we explicitly compute $\hat k$ for the cubic extended NMG model defined by eq.~\eqref{LENMG}, leading to the central charge
\begin{equation}\label{ENMGcc}
c^{\text{\tiny ENMG}} = \frac{3\ell}{2G}\sigma_4  = \frac{3\ell}{2G} \left( \sigma + \frac{1}{2 m^2\ell^2} - \frac{a}{8m^4 \ell^4} \right)\,.
\end{equation}
The central charge is proportional to the earlier defined parameter $\sigma_4$. Even though the parameter $b$ does not appear in the expression for the central charge, it does play a role in the analysis of the critical lines and points.

Appendix \ref{ExtendedCSG} deals with the parity-odd extension of $S_1$ which is denoted as $S_3$ in \eqref{S3}. Using the results derived there, we find
\begin{equation}\label{ECSGcc}
c_{3} = - \tilde{c}_{3} = \frac{3}{2\mu G} \,.
\end{equation}
This expression is equivalent to the central charge in conformal gravity; the higher-derivative terms in \eqref{S3} do not contribute to the central charge. We will address this point in more detail in section \ref{sec:ctheorem}.

% There are special points in the parameter space of ENMG where logarithmic modes appear. In that case, the central charge vanishes and the Brown-Henneaux boundary conditions can be relaxed to include a logarithmic fall-off behaviour towards the boundary of AdS$_3$. The corresponding boundary conditions are \cite{Bergshoeff:2012ev}

\subsection{Logarithmic anomalies}\label{LCFT}

The linearized analysis in section \ref{linearth} showed that the presence of massive spin-2 ghosts or tachyons in a general extended new massive gravity model cannot be avoided for non-zero FP masses. However, there are critical points where one or both of the FP masses vanish. At these points where the central charge also vanishes, new logarithmic modes appear and the linear theory is no longer diagonalisable. In that case the  gravitational theory is conjectured  to be dual to a  logarithmic conformal field theory (LCFT). % of rank 2 and 3 respectively
Knowledge of the central charge and the weights of the bulk modes is sufficient to fix the structure of the two-point functions at the critical line and at the tricritical point \cite{Grumiller:2010tj}. Here we concentrate on the left-moving sector whose  spectrum is given in the table below.
\vskip .5truecm

\begin{center}\label{hm0}\label{hleftc8}
\begin{tabular}{c| |c  }
	    & $(h, \tilde h)$ \\\hline
    $T(z)$ & $(2, 0)$ \\\hline
    $\mathcal{O}_{\pm}(z,\bar z)$ & $\left(\tfrac{3}{2} + \tfrac12 \sqrt{1+\ell^2 \cM_{\pm}^2}, -\tfrac12 + \tfrac12 \sqrt{1+\ell^2 \cM_{\pm}^2} \right)$ \\
\end{tabular}
\end{center}
\noindent {\small Table 1\ \  This table indicates the conformal weights of the operators $T(z)$ and $\mathcal{O}_{\pm}(z,\bar z)$. The expressions for $\cM_{\pm}$ can be obtained from eq.~\eqref{MM}.}
\vskip .3truecm 
Similar results hold for the right-moving sector as the two sectors are related by a parity transformation $h\leftrightarrow\tilde h$. %to compute the new anomalies of the dual LCFT
Criticality  happens whenever the conformal weights of these operators degenerate with the conformal dimension of the energy momentum tensor ($h_\pm=2$)\,\footnote{In principal one would also expect a logarithmic behavior when the two masses degenerate and hence $h_+=h_-\neq2$. This might define a LCFT with a non-zero central charge.}.  In the non-critical case, the two-point function of the left-moving components of the boundary stress tensor $T(z)$ is given by
\begin{align}
 \langle T(z) \,T(0) \rangle = \frac{c}{2 z^4} \,,
\end{align}
where $c$ is given by eq.~\eqref{ENMGcc}. If we tune $a$ to its critical value $a_{\rm crit}$, defined such that $\sigma_4=0$ and the central charge \eqref{ENMGcc} vanishes; %\cite{Grumiller:2008qz}
\begin{equation}\label{acrit}
a = a_{\rm crit} = 4 \ell^2 m^2(1 + 2 \ell^2 m^2 \sigma)\,,
\end{equation}
then one of the two masses vanishes, which we take to be $\mathcal{M}_-$, and its corresponding  boundary operator becomes the logarithmic partner of $T$. They form a Jordan cell of rank 2 with the
following two-point functions,
\begin{subequations}\begin{align}
& \langle T(z) \,T(0) \rangle = 0 \,, \\
& \langle T(z) \,\mathcal{O}_1(0) \rangle = \frac{B_1}{2 z^4} \,, \\
& \langle \mathcal{O}_1(z,\bar{z}) \,\mathcal{O}_1(0) \rangle = -\frac{B_1\,\log|z|^2}{ z^4} \,.
\end{align}\end{subequations}
The new anomaly $B_1$ can be computed through the limiting procedure of \cite{Grumiller:2010tj};
\begin{align} \label{bL}
	B_1&= \lim_{a \to a_{\rm crit}} \frac{c}{2 - h_{-}} = -\frac{24\ell}{G}\left(\sigma+\frac{1}{4m^2\ell^2}-\frac{b}{4m^4\ell^4}\right)\,.% \frac{6 B(b,m^2)}{G\ell^3m^4} = - \frac{6 b \cM_+^2}{\ell G m^4}\,.
\end{align}
% where $c_0=\frac{3\ell}{2G}$.=\frac{6\ell\Theta(a_{\rm crit})}{Gm^2}
Note that in the limit $b\to0$ and $a_{\rm crit} \to 0$ the cubic extended NMG model reduces to the critical NMG model and the result \eqref{bL} agrees with the new anomaly of NMG found in \cite{Grumiller:2009sn,Alishahiha:2010bw}.

The last case we consider is the one where $b$ also takes a critical value $b = b_{\rm crit} = \ell^2 m^2(1+4\ell^2 m^2 \sigma)$ such that $B_1=0$. This defines the tricritical point where $\sigma_4=\Theta=0$ and  both FP masses vanish.
At this point, we conjecture that the correlators are those of a rank-3 LCFT with zero central charges:
\begin{subequations}\label{2ptlcft}\begin{align}
& \langle T(z) \,T(0) \rangle = \langle T(z) \,\mathcal{O}_1(0) \rangle =0 \,, \\
& \langle T(z) \,\mathcal{O}_2(0) \rangle = \langle \mathcal{O}_1(z) \,\mathcal{O}_1(0) \rangle = \frac{B_2}{2 z^4}  \,, \label{loglog} \\
& \langle \mathcal{O}_1(z,\bar{z}) \,\mathcal{O}_2(0) \rangle = -\frac{B_2\,\log|z|^2}{z^4} \,, \\
& \langle \mathcal{O}_2(z,\bar{z}) \,\mathcal{O}_2(0) \rangle = \frac{B_2\,\log^2|z|^2}{z^4}\,.
\end{align}\end{subequations}
Here $\mathcal{O}_1(z,\bar{z})$ and $\mathcal{O}_2(z,\bar{z})$ are the two logarithmic partners of $T(z)$.
The new anomaly $B_2$ at the tricritical point is obtained via a second limit:
\begin{equation}\label{aL}
	B_2=\lim_{b \to b_{\rm crit}} \frac{B_1}{2- h_{+} }    =\frac{96\ell}{G} \left( \sigma + \frac{1}{4m^2 \ell^2} \right)\,.
\end{equation}
Note that after fixing $a$ and $b$ to their critical values, the free parameter $m^2$ is undetermined in the expression for $B_2$. This implies that tricritical  cubic extended NMG in fact has a continuous line of tricritical points dual to a family of rank-3 LCFT's with different values for the new anomaly \eqref{aL} in contrast with the PET gravity model of \cite{Bergshoeff:2012ev}.

 \subsection{Holographic $c$-theorem}
 \label{sec:ctheorem}
It is well-known that the  RG flows between fixed points in a matter theory with stress tensor $\mathcal{T}_{\mu\nu}$ coupled
to gravity and with AdS vacua can be described by a metric of the form
\bea\label{anst}
ds^2=e^{2\mathcal{A}(r)}(-dt^2+d\mathbf{x}^2)+dr^2\,.
\eea
 Assuming that the null energy condition holds for the matter sector, {\it i.e.}~$\mathcal{T}_{\mu\nu}\xi^\mu\xi^\nu\geq0$ for any null vector  $\xi$, it was shown in ref.~\cite{Freedman:1999gp} that a monotonically increasing holographic $c$-function can be found in terms of $\mathcal{A}(r)$, such that it satisfies Zamolodchikov's $c$-theorem with the radial coordinate $r$ as the measure of the energy.
The null energy condition now simplifies to
\bea
-\mathcal{T}^t_t+\mathcal{T}^r_r\geq0\,,\qquad\text{for}\qquad(\xi^t,\xi^r,\xi^x)=\left(e^{-\mathcal{A}},1,0\right)\,.
\eea
A monotonically increasing holographic $c$-function can then be obtained from
\bea\label{cprime}
c'(r)=-\frac{\mathcal{T}^t_t-\mathcal{T}^r_r}{\kappa^2\mathcal{\mathcal{A}}'^2}\geq0\,.
\eea
Assuming field equations in the bulk, $\mathcal {E}_{\mu\nu}=\kappa^2\mathcal{T}_{\mu\nu}$, the null energy condition can equivalently be written as $\mathcal {E}_{\mu\nu}\xi^\mu\xi^\nu\geq0$.  In \cite{Sinha:2010ai,Myers:2010tj} it is argued that
one way to make $c'(r)$ fulfill the  inequality \eqref{cprime} is to have $c(r)$ be only a function of $\mathcal{A}'$ which implies that
$\mathcal {E}^t_t-\mathcal {E}^r_r$ should only be a function of $\mathcal{A}'$ and $\mathcal{A}''$.  They used this logic to constrain higher-derivative interactions by demanding the presence of such a monotonically increasing function.

 Here we show that the construction \eqref{ENMGeom} is consistent with this assumption. The ansatz \eqref{anst} is conformal to AdS spacetime which is an Einstein metric and all solutions of the Einstein equations in three dimensions
 are also solutions of $C_{\mu\nu}=0$. This has the following two consequences:
 \begin{enumerate}\label{C=0}
 \item All fields which are constructed from the Cotton tensor and its derivatives are zero on the background \eqref{anst}.
 In other words, all $h$-fields and $\mathcal Df$ terms become zero on the ansatz \eqref{anst}, which means all equations \eqref{ENMGeom} reduce to a set of algebraic equations among the $f$-fields in terms of the Schouten tensor which is second order in derivatives of metric. 
 Hence the bulk field  equations involve only $\mathcal{A}'$ and $\mathcal{A}''$ by construction.
 \item Consequently, we can afford terms in the action constructed from the Cotton tensor as higher-derivative corrections without affecting
 the $c$-function. This also suggests that the only consistent way to include $\nabla R$ terms in the action is to use the Cotton tensor.
 \end{enumerate}

Now if we only focus on the bulk actions  \eqref{SNeven} and \eqref{SNodd}, this suggests that terms containing
$h$-fields and $\mathcal{D}f$-terms  do not directly contribute to one-point functions around the AdS vacuum.
The variation of the action \eqref{SNeven} around the background \eqref{anst} is only affected by $\langle \bar f\wedge \bar f\wedge \delta f\rangle$-terms because fluctuations in other terms are always
 proportional to a power of $\bar h$- or $\bar {\mathcal{D}}\bar f$-term which is zero for \eqref{anst}. In the metric formulation this means that the linearized theory around \eqref{anst} is not affected by terms 
 where graviton fluctuations are proportional to a power of the Cotton tensor which is zero for this background\footnote{
The value of the central charge is not fully determined by the equations of motion. There is always a total derivative ambiguity
which should be fixed by adding suitable boundary terms to the action and imposing suitable boundary conditions.}.
In fact, this is confirmed by direct calculation of the central charge for the first few parity-even models \eqref{NMG2} and \eqref{EENMG}, 
\bea
c_{\text{\tiny even}}=\frac{3\ell}{2G}\left(\sigma+\frac{1}{2m^2\ell^2}-\frac{a}{8m^4\ell^4}+\frac{\kappa_1}{16m^6\ell^6}+ \cdots\right)\,.
\eea
The dots refer to higher-derivative contributions to the central charge.
By the same reasoning, the variation of the action \eqref{SNodd} around the background \eqref{anst} is only affected by $\langle \bar f\wedge \bar f\wedge \delta h\rangle$-terms. But this term 
is also zero for any maximally symmetric spacetime such as AdS. In the metric formulation this is more transparent from the fact that $\bar{g}^{\mu\nu} \delta C_{\mu\nu}=0$.
So the interaction terms in the odd sector do not contribute to the central charge either. This means that the central charge in the parity-odd models is universal \eqref{ECSGcenc} and not affected by any higher-derivative term,
\bea\label{centodd}
c_{\text{\tiny odd}}=-\tilde c_{\text{\tiny odd}}=\frac{3}{2\mu G}\,.
\eea
We conclude that only interaction terms constructed solely from the Schouten tensor can contribute to the central charge. 
This is consistent with earlier studies of the holographic $c$-theorem in this context\cite{Paulos:2010ke}. Terms involving the Cotton tensor are allowed by the holographic $c$-theorem but do not contribute to the central charge ---
these terms however can contribute to the two point functions as we saw in section \ref{LCFT}. 

\section{Discussion}
In this paper, we proposed a systematic procedure of constructing higher-derivative extensions of 3D general relativity which are free of scalar ghost degrees of freedom and propagate massive spin-2 excitations.
We have used the Chern-Simons-like formulation of 3D gravity with  auxiliary form fields to find these specific scalar ghost-free combinations, which we gave explicitly up to eighth order in derivatives of the metric. These combinations turn out to be consistent with supersymmetry \cite{Bergshoeff:2014ida} and  a holographic $c$-theorem.
The number of these propagating spin-2  degrees of freedom --- some of which may be ghosts --- is determined by the number of auxiliary fields we introduce.
We considered only theories which can be written in terms of a single metric and with a finite number of ghost-free combinations --- for CS-like theories without a single-metric action see \cite{Bergshoeff:2013xma,Bergshoeff:2014pca,Afshar:prep}. 
% Here we discuss the main outcomes of this paper.
\subsubsection*{Born-Infeld gravity}
One can also construct
single-metric theories with an infinite number of ghost-free terms; as an example we consider the following  extension of NMG with a $\langle f\wedge f\wedge  f\rangle$-term in its CS-like formulation:
\begin{equation}\label{ESinha}
L = - \sigma e_a R^a + \frac{\Lambda_0}{6} \epsilon_{abc} e^a e^b e^c + h^a \cD e_a - \frac{1}{m^2} f_a\left( R^a + \frac12 \epsilon^a{}_{bc} e^b f^c + \frac{a}{6m^2} \epsilon^a{}_{bc} f^b f^c \right) \,,
\end{equation}
where $a$ is a free dimensionless parameter. The field equations obtained by varying w.r.t. $f$ are given by
\begin{equation}\label{Teom}%\label{feom}
R^a + \epsilon^a{}_{bc} e^b f^c+ \frac{a}{2m^2} \epsilon^a{}_{bc} f^b f^c = 0\,.
\end{equation}
This equation can be solved for $f$ in terms of  an infinite expansion 
\bea
f_{\mu\nu} = \sum_{n=0}^{\infty} \frac{1}{m^{2n}} f^{(n)}_{\mu\nu}
\eea
 as follows:
\begin{equation}\label{fiterative}
f^{(n+1)}_{\mu\nu} = - \frac{a}{2} \left(g_{\mu\rho}g_{\nu\sigma} - \frac12 g_{\mu\nu} g_{\rho\sigma} \right) \epsilon^{\rho\alpha\beta} \epsilon^{\sigma \gamma\delta} \sum_{k=0}^{n} f^{(k)}_{\alpha\gamma} f^{(n-k)}_{\beta\delta} \,.
\end{equation}
The starting value at order $m^0$ is $f^{(0)}_{\mu\nu} = - S_{\mu\nu}$. Having found the solution for $f_{\mu\nu}$, we can go to the metric formulation by plugging the solution of \eqref{fiterative} back into the Lagrangian. The result can be written as
\begin{equation}\label{LESinha}
\cL = \frac{1}{2} \left\{ \sigma R - 2 \Lambda_0 - \frac{2}{m^2} \left[ \frac{2}{3} f_{\mu\nu} G^{\mu\nu} + \frac16 \left( f_{\mu\nu} f^{\mu\nu} - f^2 \right) \right] \right\}\,.
\end{equation}
Here $ f = f_{\mu\nu} g^{\mu\nu}$ and $f_{\mu\nu}$ is given in terms of the coefficients $f^{(n)}_{\mu\nu}$ in eq.~\eqref{fiterative}. Explicitly, up to order $1/m^6$ we have checked that the scalar ghost free combinations in \eqref{EENMG} are recovered  with the Cotton tensor set to zero. When $a=\sigma$ these are the  same leading terms that occur in the expansion of the Born-Infeld extension of NMG \cite{Gullu:2010pc,Gullu:2010st} --- see \cite{Bergshoeff:2014ida} for a supersymmetric version.
Our construction \eqref{LESinha} compares nicely with an earlier proposal based on limits of a class of bimetric theories \cite{Paulos:2012xe}. 
The linearized spectrum of this model includes only one massive graviton.
Using the prescription explained in section \ref{sec:cc} and appendix \ref{can} we obtain the following expression for the central charge of the model:
\begin{equation}
 c=\frac{3\ell}{2G}\left[\sigma+\frac{1}{a}\left(-1+\sqrt{1+\frac{a}{m^2\ell^2}}\,\right)\right]\,.
\end{equation}
This coincides with the central charge computed in \cite{Gullu:2010st}, when $a=\sigma$.  As shown in \cite{Jatkar:2011ue,Sen:2012fc} for the choice of parameters where the central charge becomes zero, 3D Born-Infeld gravity arises as a suitable counterterm for gravity in AdS$_4$. %If $a>0$, the value of the central charge can increase arbitrarily, but when $a<0$, it is bounded.

\subsubsection*{Holographic $c$-theorem}
Considering these higher-derivative models as toy models for exploring the role of
higher-derivatives in holography, we observed that higher-derivative theories can accommodate terms involving $\nabla R$ which are introduced via the Cotton tensor in special combinations --- see for instance \eqref{EENMG} and \eqref{EECSG}. They are fully consistent with a holographic $c$-theorem. 
This feature can easily be generalized to higher dimensions by using the higher-dimensional Cotton and Schouten tensors:
\bea 
 C_{\mu\nu\lambda} = (D-2)\left(\nabla_{\lambda} S_{\mu\nu} - \nabla_{\nu} S_{\mu\lambda} \right) \quad\text{and}\quad S_{\mu\nu}= \tfrac{1}{(D-2)} \left( R_{\mu\nu}-\tfrac{1}{2(D-1)}R\,g_{\mu\nu} \right) \,.
\eea
In a sense, these tensors seem to be the right basis for studying higher-derivatives in the context of holography.
Using the first order formulation, we particularly showed that the Einstein equations of higher-derivative theories in three dimensions, evaluated on the background \eqref{anst}, are always second order in derivatives. 

Terms containing the Cotton tensor,  do not contribute to the AdS  one-point functions including the central charge. 
This covers a broader class of higher-derivative theories admitting a holographic $c$-theorem than the class of theories considered in \cite{Myers:2010tj,Sinha:2010ai,Paulos:2010ke}, which only included higher-curvature terms containing $R^n$ tensors.

Since the presence of these higher-derivative terms leads to tachyons or ghosts, this observation confirms the 
conclusions that some `unphysical' models with non-unitary operators still  satisfy a holographic $c$-theorem \cite{Myers:2010tj}. 
% Demanding unitarity of the boundary theory however should guarantee that the bulk theory obeys holographic $c$-theorem.
%  unitarity of the boundary theory along the
% RG ows also guarantees that the theory obeys a holographic c-theorem.(NOT)
% (higher-curvature vs higher-derivative) We made this observation also for higher-derivative terms! Cotton tensors do contribute to linearized eom
% such modes will not (WILL) be associated with the
% appearance of new operators in the boundary theory

\subsubsection*{Unitarity}
The problem of non-unitarity is generic in higher-derivative models including the ones we consider here. 
The linear spectrum of these higher-derivative  theories generically propagates massive spin-2 modes. However, because of the higher-derivative nature of the theory, some of these massive modes are inherently unstable; this result is compatible with earlier higher-derivative extensions of general relativity, see for instance \cite{Nutma:2012ss,Bergshoeff:2012ev}.
Due to the instability of the linearized massive modes, the applicability of these models may be limited to special, critical points in their parameter space; 
points where the linearized equations are only second order in derivatives and the massive modes disappear %and spots a unitary point and a unitarity dual CFT, if any. 
or points where they become massless and are replaced by log modes with logarithmic fall-off behavior towards the AdS boundary. At these latter points, the dual CFT is expected to be a logarithmic CFT. Although LCFT's are non-unitary, they have applications in statistical physics --- see \cite{Grumiller:2013at,Cardy:2013rqg} for recent reviews.

The higher-derivative nature of the models we constructed in this work, exhibits the same ``bulk-boundary unitarity problem"  that is inherent to higher-derivative extensions of general relativity. This problem refers to the impossibility to obtain a positive boundary central charge (or black hole charge) for a region in parameter space that has  well-behaved bulk spin-2 modes. Recently, a different CS-like theory for 3D gravity was introduced which resolves this problem.
This model was called zwei-dreibein gravity (ZDG) \cite{Bergshoeff:2013xma}. The resolution of the bulk-boundary conflict stems from the fact that the parameter region of ZDG is large enough to include a  well-behaved region as
far as the  sign of the central charge, the kinetic terms and the mass terms of the theory are concerned. This enhancement in parameters has the consequence that the ZDG  action cannot be written in terms of (higher-derivatives of) a single metric.\footnote{This is not true for the equations of motion, which feature an infinite expansion of higher-derivative terms \cite{Bergshoeff:2014eca}.} 
We expect that the bulk-boundary clash in the higher-derivative extensions of general relativity presented in this paper can similarly be resolved by considering a ZDG-like extension, generically called ``viel-dreibein gravity'', that involves more than 4 Lorentz vector-valued one-form fields \cite{Afshar:prep}.

Here we have only concentrated on AdS holography. These models can accommodate  non-AdS spacetimes due to a large parameter space and might be unitary on these backgrounds.
Specifically, there is a quantization preference in three dimensions in terms of unitarity.  Unitary quantization of parity-odd theories seems to prefer asymptotically flat spacetimes, while AdS asymptotics are good for quantization of parity-even theories. 
Flat boundary conditions at null infinity in three dimensional gravity lead to a centrally extended BMS$_3$ algebra as asymptotic symmetry algebra \cite{Barnich:2006av} --- for a new derivation see \cite{Afshar:2013bla}. The non-zero commutators are,
\begin{align}
 [L_m,L_n]&=(m-n)L_{m+n}+\frac{c_L}{12}m(m^2-1)\delta_{m+n,0}\,,\\
 [L_m,M_n]&=(m-n) M_{m+n}+\frac{ c_M}{12}m(m^2-1)\delta_{m+n,0}\,,
\end{align}
with Virasoro generators $L_n$ and supertranslations $M_n$.  A unitary theory in flat space should have $c_M=0$. Then one is left with a single copy of the Virasoro algebra. This happens for free
for all parity-odd gravity theories like conformal gravity \cite{Bagchi:2012yk,Afshar:2013bla}  --- see \cite{Afshar:2014rwa} for a recent review. For all parity-odd higher-derivative gravity theories 
that we considered here the value of the AdS central charges are the same \eqref{centodd}, therefore they lead to the same asymptotically flat symmetries with 
\bea
c_L=\frac{3}{\mu G}\,,\qquad\qquad c_M=0\,.
\eea
This is a necessary condition for unitarity but not sufficient. The higher-derivative parity-odd theories with flat boundary conditions might suffer from the same pathologies as in the parity-even ones with AdS boundary conditions; the massive  modes can propagate negative norm states.

Another interesting development would be to construct a holographic dictionary for the class of Chern-Simons-like models with local bulk degrees of freedom. 
% In this work we only considered the asymptotic symmetry algebra, which contains information about the two-point functions of the stress-tensor of the dual CFT. However, as we remarked in section \ref{Holography},  the massive modes in the bulk correspond to massive operators in the boundary CFT. At present, it is only known how to compute their two-point functions in a second order formalism  through a holographic renormalization. It would be very interesting to understand how this information can be obtained directly from the CS-like formulation of 3D gravity. 
This could then  directly be applied to 
the models we considered in this paper and to 
a variety of other CS-like models, such as ZDG and the recently introduced minimal massive gravity extension of TMG  \cite{Bergshoeff:2014pca}.

\section*{Acknowledgements}
The authors wish to thank Mehmet Ozkan for helpful discussions. We thank Daniel Grumiller, Alasdair Routh, Paul Townsend and Thomas Zojer for their useful comments on the draft. H.~R.~A. and W.~M. are supported by the Dutch stichting voor Fundamenteel Onderzoek der Materie (FOM).

\appendix

\section{Charges in Chern-Simons-like theories}\label{can}
In this appendix we discuss the identification of the first class constraints of the CS-like theories and the corresponding boundary charges, which is of use in section \ref{Holography} when we discuss the AdS holography for the higher-derivative gravity models we consider in this work.

\subsection{Canonical analysis}

The advantage of working with a Chern-Simons-like formulation of 3D higher-derivative gravity models is appreciated mostly when analyzing these theories in a Hamiltonian form and when computing their asymptotic symmetry algebra. All of the models we consider in this work belong to the following class of theories that are defined by a set of $N$ Lorentz-valued one-form fields with an action  given by \cite{Hohm:2012vh}\,\footnote{Here we use a notation where  wedge products between forms and Lorentz indices $a,b,\cdots$ are suppressed.
The dots and crosses indicate contractions with $\eta_{ab}$ and $\epsilon_{abc}$ respectively.}
\begin{equation}\label{Lgeneral}
S = \frac{1}{2\kappa^2} \int_{\Sigma\times\mathbb{R}}g_{rs}  a^r \cdot da^s + \tfrac13 f_{rst} a^r \cdot (a^s \times a^t)\,.
\end{equation}
Here $g_{rs}$ is a symmetric, constant and invertible metric on the flavor space which can be used to raise and lower flavor indices. The $f_{rst}$ are totally symmetric structure constants; the theory is pure Chern-Simons when the expressions $\epsilon^a{}_{bc}f^r{}_{st}$ are the structure constants and $g_{rs}\eta_{ab}$ is a non-degenerate bilinear form of a Lie algebra.

The Hamiltonian formalism for this class of models was performed in \cite{Hohm:2012vh,Bergshoeff:2014bia}. Here, we
 review  some of the results of these references  in order to fix the notation and derive the asymptotic symmetry algebra. After a space-time decomposition of the $N$ Lorentz-valued one-forms, the time components of the fields $a_0^{r}$ are Lagrange multipliers  and $a_i^r$ are dynamical fields satisfying the following Poisson brackets,
 \bea
 \{a_{i\,a}^r(x),a_{j\,b}^s(y)\}=\kappa^2\,\epsilon_{ij}g^{rs}\eta_{ab}\delta^2(x,y)\,.
 \eea
The  Lagrange multipliers induce a set of $3N$ primary constraints in the Hamiltonian analysis of the theory \cite{Bergshoeff:2014bia};
\begin{equation}\label{constraints}
\phi_r = \frac{1}{\kappa^2}\varepsilon^{ij} \left(g_{rs} \partial_i a_j^{s} + \tfrac12 f_{rst} \left( a_{i}^s \times a_{j}^t \right) \right)\,.
\end{equation}

 It is convenient to define the ``smeared'' functions $\phi_r[\xi^r]$ associated to the constraint functions \eqref{constraints} by integrating them against a test function $\xi^r(x)$ as follows:
\begin{equation}\label{phi}
\phi[\xi]=\sum_r\phi_r[\xi^r] = \int_{\Sigma} d^2 x \; \xi^r(x) \cdot\phi_r (x) \,.
\end{equation}
Here $\Sigma$ is a space-like hypersurface. In general, the variation of $\phi_r[\xi^r]$ may lead to non-zero boundary terms. Varying the expression \eqref{phi} for $\phi[\xi]$ with respect to the fields $a_{i}{}^s$ gives
\begin{equation}
\delta \phi [\xi] = \int_{\Sigma} d^2 x \; \xi^{r}\cdot \frac{\delta \phi_r}{\delta a_{i}^{s}}\cdot \delta a_{i}^{s} + \int_{\partial \Sigma} dx \; B[\xi, a,\delta a]\,.
\end{equation}
The boundary terms in this expression could lead to delta-function singularities in the Poisson brackets of the constraint functions. To remove these, we choose boundary conditions that make $B$ a total variation
\begin{equation}
\int dx \; B[\xi,a, \delta a] = - \delta Q[\xi, a]\,.
\end{equation}
We can then define an improved set of constraint functions via
\begin{equation}\label{varphi}
\varphi[\xi] = \phi[\xi] + Q[\xi, a]\,.
\end{equation}
These new functions  will now have a well-defined variation without boundary terms. In our case, using eq.~\eqref{constraints}, we find
\begin{equation}\label{gen_varbc}
\delta Q = - \frac{1}{\kappa^2}\int_{\partial \Sigma} dx^i\, g_{rs}\, \xi^r \cdot\delta a_{i}^{s}\,.
\end{equation}
The Poisson brackets of the constraints were computed in \cite{Hohm:2012vh,Bergshoeff:2014bia}. They are given by
\begin{align} \label{gen_poissonbr}
\left\{ \varphi(\xi) , \varphi(\eta)  \right\}_{\rm P.B.} = & \; \phi([\xi, \eta]) + \frac{1}{\kappa^2}\int_{\Sigma} d^2x \; \xi^r_a \eta^s_b \, \cP_{rs}^{ab}
\nonumber \\
& - \frac{1}{\kappa^2}\int_{\partial \Sigma} dx^i \; \xi^r \cdot \left(g_{rs}  \partial_i \eta^s + f_{rst} (a_{i}{}^s \times \eta^t)   \right)\,,
\end{align}
where we have defined
\begin{align}
\cP_{rs}^{ab} & = f^t{}_{q[r} f_{s] pt} \eta^{ab} \Delta^{pq}  +  2f^t{}_{r[s} f_{q]pt} (V^{ab})^{pq}\,, \label{Pmat_def} \\[.2truecm]
V_{ab}^{pq} & =  \varepsilon^{ij} a^p_{i\, a} a^q_{j\, b}\,, \hskip 1truecm \Delta^{pq} = \varepsilon^{ij} a_i^p \cdot a_j^q\,,\hskip 1truecm
[\xi ,\eta]^t  = f_{rs}{}^{t}  \xi^r\times \eta^s\,.
\end{align}
A detailed analysis of how to deal with the secondary constraints in this type of theories was presented in \cite{Bergshoeff:2014bia}.
It suffices to state here that the secondary constraints derived in section \ref{counting} remove the $\Delta^{pq}$-term in the matrix $\cP$ of Poisson brackets defined  in eq.~\eqref{Pmat_def}.
For the purpose of this paper, it is sufficient to focus only on the algebra of constraint functions when adapting AdS (or Brown-Henneaux) boundary conditions. In the next section we will discuss the identification of the first class constraint functions.

\subsection{First class constraint functions}\label{fcc}
In contrast to the pure Chern-Simons gauge theories, not all constraint functions \eqref{constraints} in the Chern-Simons-like models
are first class. In order to properly analyze the asymptotic symmetries, we should look at the algebra of first class constraint functions which generate the gauge symmetries. Hence, we first need to identify  which (combination of) constraint functions generate the gauge symmetries of the theory.

The general CS-like theory defined by eq.~\eqref{Lgeneral} is manifestly diffeomorphism invariant. In addition, the specific CS-like models of our interest are also manifestly invariant under local Lorentz transformations. All models defined by the actions \eqref{SNeven} and \eqref{SNodd} contain a (dualized) spin-connection $\omega$, which only appears in terms of the dualized curvature two-form $R=\cD\omega$ or via a Lorentz-Chern-Simons term in the action. Moreover, all derivatives of the other one-form fields $a^{r}$ %with $r \neq \omega$ 
appear as covariant derivatives $\cD a^{r}$. Translated to components of the flavor space metric $g_{rs}$ and $f$-tensor $f_{rst}$ this assumption is equivalent to the following statement
\begin{itemize}
\item[] {\it For every element of $g_{rs}$  there is a corresponding $f_{rs \omega}$ such that: $f_{rs \omega}=g_{rs}$.}\;\;\;\;\eqnum\label{omega_assumpt}
\end{itemize}
Equipped with this assumption we should expect the CS-like models defined by \eqref{Lgeneral} to have at least six gauge symmetries, corresponding to three diffeomorphism and three local Lorentz transformations.

To identify the constraint functions which generate these symmetries, it is instructive to look at the Poisson brackets of the gauge transformations with the dynamical components of the theory. In full generality (but omitting boundary terms 
at this point), they can be computed using the general formulas \eqref{constraints} and \eqref{phi} 
with the following result:
\begin{equation}\label{transgen}
\{ \phi[\xi], a_i^{r}\} = \partial_i \xi^{r} + f^{r}{}_{st} a_{i}^{s}\times \xi^{t}\,.
\end{equation}
From this result, we can deduce that a local Lorentz transformation with the gauge parameter $\tau$ is generated by the constraint function
\begin{equation}\label{phiLL}
\phi_{\rm LL}[\tau] \equiv \phi_\omega[\xi^\omega] \ \  \text{ with }\ \  \xi^{\omega } = \tau \,.%\ \ \text{ and }\ \ \xi^{r\,a} = 0\, \text{ for } r \neq \omega\,.
\end{equation}
With this identification  we recover the usual transformation properties under local Lorentz transformations from \eqref{transgen}:
\begin{equation}
\begin{split}
\delta_{\tau}\omega_{i} = & \{\phi_{\rm LL}[\tau] , \omega_{i}\} = \partial_i \tau + \omega_i\times \tau\,=\cD_i\tau\,, \\
\delta_{\tau}a_{i}^{r} = & \{\phi_{\rm LL}[\tau] , a_i^{r} \} =  a_i^{r} \times\tau\,, \;\qquad r\neq\omega\,,\\
\end{split}
\end{equation}
where we have used the fact that by the assumption \eqref{omega_assumpt} we may write $f^r{}_{s \omega} \equiv g^{rp}f_{ps \omega} = g^{rp}g_{ps} = \delta^r_s$.

In Chern-Simons gauge theories, diffeomorphisms are generated by an appropriate combination of constraint functions with parameters proportional to the fields \cite{Witten:1988hc}, on-shell. The same is true for
 the general CS-like theory. Let us define
\begin{equation}\label{phidiff}
\phi_{\rm diff}[\zeta] \equiv \sum_r\phi_r[\xi^r]\quad\text{with}\quad \xi^r=a_{\mu}^r \zeta^{\mu}\,.
\end{equation}
Then, by equation \eqref{transgen} we find that
\begin{equation}\label{deltadiff}
\begin{split}
\delta_{\zeta}a_{i}^{r} = \{\phi_{\rm diff}[\zeta], a_{i}^{r} \} = \cL_{\zeta} a_{i}^{r} + & \, \zeta^\mu \left( \partial_i a_\mu^{r} - \partial_{\mu} a_i^{r} + f^r{}_{st}a_i^{s}\times a_{\mu}^{t} \right) \,.
% + & \, \zeta^0 \left( \partial_i a_0^{r} - \partial_{0} a_i^{r} +  f^r{}_{st}a_i^{s}\times a_{0}^{t} \right)\,.
\end{split}
\end{equation}
Here $\cL_{\zeta}$ is the Lie derivative with respect to the vector field $\zeta^{\mu}$. The expressions in the parentheses are equivalent to the equations of motion of the general model \eqref{Lgeneral} after a space-time decomposition. Hence, on-shell we have 
% \begin{equation}
% \delta_{\rm diff}a_{i}^{r\,a} = \{\phi_{\rm diff}[\zeta], a_{i}^{r\,a} \} = \cL_{\zeta} a_{i}^{r\,a}\,.
% \end{equation}
identified the constraint functions which give the correct transformation rules on the dynamical variables of the theory.

\subsection{Boundary charges}%Algebra of first-class constraints and b
To proceed with an analysis of the asymptotic symmetry group for AdS boundary conditions, we would like to write the algebra of first class constraint functions in a basis of mutually commuting $SL(2,\mathbb{R})$ generators. This is  possible on the AdS background as the generators of gauge symmetries should respect the isometries of the AdS vacuum solution. %At the boundary of an asymptotically AdS spacetime, the AdS background identities hold and hence the new basis of first class constraint functions is also valid there. 
In all parity preserving models we can define such a basis as
\begin{equation}\label{Lpm}
J_{\pm} [\zeta] = \tfrac12\left(\phi_{\rm LL}[e_{\mu}\zeta^{\mu}]\pm\ell \,\phi_{\rm diff}'[\zeta] \right)\,,
\end{equation}
where $\phi_{\rm diff}'[\zeta] = \phi_{\rm diff}[\zeta] - \phi_{\rm LL} [\omega_{\mu}\zeta^{\mu}]$.
% These generators are defined such that
% \begin{equation}\label{LplusLminus}
% \{L_+[\xi],L_{-}[\eta]\} = 0
% \end{equation}

At this point one should reinstate the boundary terms introduced in  eq.~\eqref{varphi}, and investigate the Poisson bracket algebra of the generators \eqref{Lpm} subject to asymptotically AdS (or Brown-Henneaux) boundary conditions. The improved  differentiable  generators are then,
\bea\label{improvedJ}
\mathcal{J}_{\pm} [\zeta]=J_{\pm} [\zeta]+Q_{\pm}[\zeta]\,.
\eea
% The algebra of the Poisson brackets  becomes,
% \begin{equation}\label{Lpmbrackets}
% \begin{split}
% \{ \mathcal{J}_{\pm} [\zeta_1], \mathcal{J}_{\pm} [\zeta_2] \} &=  J_{\pm}[\zeta_3] +\tfrac12\left(\delta_{\zeta_2} Q_{\pm}[\zeta_1]-\delta_{\zeta_1} Q_{\pm}[\zeta_2]\right)\,,\\
% \qquad \{\mathcal{J}_+[\zeta_1],\mathcal{J}_{-}[\zeta_2]\} &= 0\,.
% \end{split}
% \end{equation}
% where $\zeta_3=\zeta_1\cdot\partial\zeta_2-\zeta_2\cdot\partial\zeta_1$ is the Lie bracket of $\zeta_1$ and $\zeta_2$ \cite{Afshar:2013bla}. The variation of the boundary terms on the right hand side can be computed from \eqref{gen_varbc} and \eqref{deltadiff}.

Let us first comment that, quite generally, by the fact that the auxiliary fields are symmetric (see eq.~\eqref{fhantisym}), the gauge parameters for diffeomorphisms $\xi^r = a_{\mu}{}^r \zeta^{\mu}$ satisfy
\begin{equation}
\begin{split}
 e_{\mu}\cdot \xi^{f_I} = e_{\mu}\cdot  f_{I\,\nu} \zeta^{\nu} = f_{I\,\mu}\cdot  e_{\nu} \zeta^{\nu} = f_{I\,\mu}\cdot  \xi^e\,,  \qquad e_{\mu}\cdot  \xi^{h_I} = h_{I\,\mu}\cdot  \xi^e\,.
\end{split}
\end{equation}
Moreover,  on the AdS background, since the auxiliary fields $f_I$ are all proportional to the AdS dreibein and the auxiliary fields $h_I$ vanish, we have
\begin{equation}\label{gpbkgd}
\xi^{f_I} = \bar{f}_I \xi^e \,, \qquad \xi^{h_I} = 0\,,
\end{equation}
where $\bar{f}_I$ is the constant background value of the auxiliary fields, {\it i.e.}~$\bar{f}_I{}^a = \bar{f}_I \bar{e}^a$ on the AdS background. The values for $\bar{f}_1$ and $\bar{f}_2$ can be read from eq.~\eqref{fluctuations}. This allows us to express all the gauge parameters occurring  in $\phi_{\rm diff}'[\zeta]$ in terms of $\xi^e=e_{\mu}\zeta^\mu$. Using these relations, we can compute the variation of the boundary terms in \eqref{improvedJ}  from \eqref{gen_varbc};
\begin{equation}
\begin{split}\label{Qpm}
\delta Q_{\pm}[\xi^e_\pm] = & - \frac{1}{2\kappa^2}\int_{\partial \Sigma} dx^i \; \left(g_{\omega s}\, \xi^e_\pm \cdot\delta a_{i}^{s}  \pm\ell\, g_{rs}\,\xi^r_\pm \cdot\delta a_{i}^{s}\right)  \,,\\
=& - \frac{1}{2\kappa^2}\int_{\partial \Sigma} dx^i \;\xi^e_\pm \cdot\left[ g_{\omega s} \,\delta a_{i}^{s}+ \cdots \pm\ell \left(  g_{es} + \bar{f}_1 g_{f_1s} + \bar{f}_2 g_{f_2s}\right) \delta a_{i}^{s} \right]\,,
\end{split}
\end{equation}
where in the first line the sum over $r$ does not include $\omega$ and in the second line we have used eq.~\eqref{gpbkgd}. In general, after plugging in the explicit flavor space metric and AdS background values of the fields, the result may be written as
\begin{equation}\label{Qeff}
\delta Q_{\pm}[\xi^{\pm}] =  \frac{\hat k}{2\pi} \int_{\partial \Sigma} dx^i \; \xi_a^{\pm} \left( \delta \omega_i{}^a \pm \frac{1}{\ell} \delta e_i{}^a \right)\,.
\end{equation}
where $\hat k$ is an effective dimensionless coupling determined by the elements of $g_{rs}$ and the $\bar{f}_I$'s. We have also distinguished the gauge parameters for the left and right moving sectors explicitly;
\bea
\xi^{\pm}=\pm\frac{1}{2}\xi^e_\pm\,.
\eea

After imposing suitable boundary conditions and restricting $\xi^{\pm}$ to the set of symmetry transformations which preserve these boundary conditions it is possible to integrate this expression to obtain the global conserved charges of the theory. This is done explicitly in section \ref{sec:cc} for Brown-Henneaux boundary conditions. The effective coupling $\hat{k}$ is related to the central charge as $c = \tilde{c} = 6 \hat{k}$ for the parity-even models and $c = - \tilde{c} = 6 \hat{k}$ for the parity-odd models.

As an example of these general considerations, let us compute the effective coupling $\hat{k}$ for ENMG explicitly. The parity-odd theory defined by $S_3$ will be treated in full detail in appendix \ref{ExtendedCSG}.
% It is convenient to first define the constraint functions
% \begin{equation}
% \phi_t[\xi^t] = \int d^2x \; \xi^t_a \phi^a_t \,,
% \end{equation}
% where it is understood here that there is {\it no sum over $t$}. Our aim is first to find the generators \eqref{Lpm} such that \eqref{LplusLminus} holds and consequently compute the corresponding boundary charges \eqref{Qpm}.
% 
Using the above relations and the specific values of $g_{rs}$ and $f_{rst}$ for ENMG, which may be read off from eq.~\eqref{LENMG}, 
we find that the variation of the conserved charges are given by \eqref{Qeff} with:
\begin{equation}\label{kENMG}
\hat k^{\text{\tiny ENMG}} = \frac{\ell}{4G} \left(\sigma + \frac{1}{2m^2 \ell^2} - \frac{a}{8m^4 \ell^4}\right)\,.
\end{equation}

\section{Extended gravitational Chern-Simons term}\label{ExtendedCSG}

In this appendix we consider the class of parity-odd theories given in eq.~\eqref{S3}. Applying the same
procedure as in the parity-even case the following expression for the  Lagrangian 3-form  that
describes the leading five derivative  extension in the odd sector can be derived:
\begin{equation}\label{LECSG}
\begin{split}
 L_3 =   \frac{1}{2\mu}\omega_a \left(d \omega^a + \frac{1}{3} \epsilon_{abc} \omega^b \omega^c\right) +\frac{1}{\mu^3}\Big[&e_a \cD f_2{}^a+ h_{1\,a} \left( R^a + \epsilon^{abc}e_b f_{1\,c} \right)\\& + \frac{\alpha}{2}\,  f_{1\,a}\cD f_1{}^a  \Big]  \,.
\end{split}
\end{equation}
Here, $\alpha $ is an arbitrary dimensionless parameter.
The equations of motion for this Lagrangian, obtained by varying with respect to $f_2{}^a$, $h_1{}^a$, $f_1{}^a$, $\omega^a$ and $e^a$, respectively, are given by
\begin{equation}\label{ECSGeom2}
\begin{split}
& \cD e^a = 0 \,, \\
& R^a + \epsilon^a{}_{bc} e^b f_1{}^c = 0\,, \\
& \alpha \cD f_1{}^a +  \epsilon^a{}_{bc} e^b h_1{}^c = 0 \,, \\
&\cD h_1{}^a     + \tfrac{1}{2} \epsilon^a{}_{bc}\left(\alpha \,f_1{}^b f_1{}^c + 2\epsilon^a{}_{bc}e^b f_2{}^c - 2\mu^2 \sigma \epsilon^a{}_{bc} e^b f_1{}^c\right) = 0\,, \\
& \cD f_2{}^a + \epsilon^a{}_{bc} f_1{}^b h_1{}^c = 0 \,.
\end{split}
\end{equation}
 Upon acting on the equations of motion with an exterior derivative and performing some algebra, one can derive that the auxiliary fields are symmetric
\begin{equation}
f_{1\,[\mu\nu]} = h_{1\,[\mu\nu]} = f_{2\,[\mu\nu]} = 0\,.
\end{equation}
We can solve them in turns of derivatives of the dreibein. Explicitly, one finds the following expressions:
\begin{align}
 f_{1\,\mu\nu} = - S_{\mu\nu}\,, \quad h_{1\,\mu\nu} = \alpha C_{\mu\nu} \,, \quad
 f_{2\, \mu\nu} = - \alpha \,D_{\mu\nu} + \alpha\left( P_{\mu\nu}-\tfrac{1}{4}P\,g_{\mu\nu}  \right) - \mu^2 S_{\mu\nu}  \,,
\end{align}
where $P_{\mu\nu}$ and $D_{\mu\nu}$ are defined in \eqref{DPEQten}.
Substituting these solutions back into the action leads to the following extended  Lagrangian density: %Chern-Simons gravity\footnote{We dropped the term ``conformal'' since this symmetry is broken now.}  (ECSG)
\begin{align} \label{ECSG}
\mathcal{L}_3&=\mathcal{L}_1+\frac{\alpha}{\mu^3}S_{\mu\nu}C^{\mu\nu}\nn\\
&=\frac{1}{\mu}\epsilon^{\mu\nu\lambda}\left\{\Gamma_{\mu\sigma}^\rho\partial_\nu\Gamma^\sigma_{\lambda\rho}+\frac{2}{3}\Gamma^\rho_{\mu\sigma}\Gamma^\sigma_{\nu\tau}\Gamma^\tau_{\lambda\rho}+\frac{\alpha}{\mu^2}\, R_\mu{}^\sigma\nabla_\nu R_{\sigma\lambda} \right\}\,.
\end{align}
% where $a$ and $b$ are arbitrary parameters and
Applying the same procedure one step further we find the following  seven derivative  Lagrangian density:
\begin{align}\label{EECSG}
\mathcal{L}_5=\mathcal{L}_{3}+\frac{1}{\mu^5}\bigg\{\beta_1P_{\mu\nu}C^{\mu\nu}+\beta_2D_{\mu\nu} C^{\mu\nu}\bigg\}\,,
\end{align}
where $\beta_1$ and $\beta_2$ are two dimensionless parameters and $P_{\mu\nu}$ and $D_{\mu\nu}$ are defined in \eqref{DPEQten}.

\subsection{Linearization}
To obtain the quadratic Lagrangian 
we substitute the fluctuations \eqref{fluctuations} into the action \eqref{LECSG} with $m\rightarrow\mu$ and $a\rightarrow \alpha$. After making the  field redefinitions
\bea\label{fredef}
k_2\rightarrow k_2+(\mu^2+\tfrac{\alpha\Lambda}{2})k_1-\tfrac{\alpha\Lambda^2}{8}k\qquad\text{and}\qquad k_1\rightarrow k_1-\tfrac{\Lambda}{2}k\,,
\eea
we obtain the following  quadratic Lagrangian for $S_3$:
\begin{equation}\label{quadL3}
\begin{split}
L_{3}^{(2)} =  L_{1}^{(2)}+\frac{1}{\mu^3} \bigg\{& k_{2\,a}  \left(  \bar{\cD} k^a + \epsilon^a{}_{bc} \bar{e}^b v^c \right)  +  v_{1\,a} \left( \bar{\cD} v^a - \Lambda \epsilon^a{}_{bc} \bar{e}^b k^c +  \epsilon^a{}_{bc} \bar{e}^b k_1^c \right)\\&+\frac{\alpha}{2}k_1\bar{\cD}k_1\bigg\}\,.
\end{split}
\end{equation}
Upon eliminating the auxiliary fields by their equations of motion, we find the five derivative  Lagrangian density
\begin{equation}
\cL_3^{(2)} =  - \frac{1}{\mu}\epsilon^{\mu\alpha}{}_{\rho} \bar{\nabla}_{\alpha} k^{\rho\nu} \cG_{\mu\nu}(k) + \frac{\alpha}{\mu^3} \epsilon^{\mu\alpha}{}_{\rho} \bar{\nabla}_{\alpha} \cG^{\rho\nu}(k) \cG_{\mu\nu}(k)\,,
\end{equation}
where $\cG_{\mu\nu}$ is defined in \eqref{linG}.
The linearized equations of motion for the action corresponding to this Lagrangian density can be written as:
\begin{equation}
(\cD^0 \cD^M \tilde{\cD}^M \cD^L \cD^R k)_{\mu\nu} = 0\,,
\end{equation}
where %\bea\label{DLDRopt}
$(\cD^{L/R})_{\mu}^{\rho} = \delta_{\mu}^{\rho} \pm  \ell \epsilon_{\mu}{}^{\alpha\rho} \bar{\nabla}_{\alpha}
 $ 
 and 
the differential operator $\cD^{0}$ is defined in \eqref{DOopt} while $\cD^{M}$ and $\tilde{\cD}^M$ are defined as in \eqref{DMopt}
% :
% \begin{align}
% (\cD^M)_{\mu}^{\rho} = \delta_{\mu}^{\rho} + \frac{1}{M} \epsilon_{\mu}{}^{\alpha\rho} \bar{\nabla}_{\alpha} \,, &&
% (\tilde{\cD}^M)_{\mu}^{\rho} = \delta_{\mu}^{\rho} - \frac{1}{M} \epsilon_{\mu}{}^{\alpha\rho} \bar{\nabla}_{\alpha} \,,
% \end{align}
% and 
 with the mass parameter $M$ given by:
\begin{equation}
M =  \sqrt{ \cM^2 -  \Lambda } =
\sqrt{\frac{1}{\ell^2} -\frac{\mu^2}{\alpha}}\,.
\end{equation}
The linear theory hence describes a partially massless mode, and two helicity-$\pm 2$ massive modes, with a Fierz-Pauli mass $\cM^2 = - \mu^2/ \alpha$. Note that the conformal symmetry is broken due to the additional interaction term. The theory hence propagates three degrees of freedom, corresponding to the two helicity states of the massive mode and the partially massless mode.

In accordance with what is expected from the linear spectrum, it is possible to diagonalize the quadratic Lagrangian. After making the appropriate field redefinitions the Lagrangian density \eqref{quadL3} can be written as:
\begin{align}
L_3^{(2)} = & \; \frac{1}{\mu} \left( k_{L}^a \bar{\cD} k_{L\,a} + \frac{1}{\ell} \epsilon_{abc} \bar{e}^a k_{L}^b k_{L}^c \right)
+ \frac{1}{\mu} \left( k_{R}^a \bar{\cD} k_{R\,a} - \frac{1}{\ell} \epsilon_{abc} \bar{e}^a k_{R}{}^b k_{R}^c \right) \nonumber \\
& + \frac{(\alpha - \ell^2 \mu^2)}{2 \mu} k_{0}^a \bar{\cD} k_{0\,a}  - \frac{1}{\mu}\left( k_{M_+}^a \bar{\cD} k_{M_+\,a} + M \epsilon_{abc} \bar{e}^a k_{M_+}^b k_{M_+}^c \right)  \\
& - \frac{1}{\mu} \left( k_{M_-}^a \bar{\cD} k_{M_-\,a} - M \epsilon_{abc} \bar{e}^a k_{M_-}^b k_{M_-}^c \right)\,, \nonumber
\end{align}
where we have  assumed that $\mu^2\ell^2 \neq \alpha$. The modes indexed by $L/R/0$ or $M_{+/-}$ are annihilated by $\cD^{L/R/0}$ and $\cD^{M}/\tilde {\mathcal{D}}^{M}$ respectively. The point $\alpha = \ell^2 \mu^2$ corresponds to a special case in the linear spectrum where the massive modes become partially massless and degenerate with the partially massless mode $k_0$.
Note that  there is no (finite) parameter choice possible where the massive mode degenerates with the massless mode and the massive and the massless sectors come with opposite signs.

\subsection{Central charge}
The extended Lagrangian \eqref{LECSG} fits the general model \eqref{Lgeneral} with flavor space metric and structure constants given by
\begin{align}
g_{\omega\omega} = \frac{1}{\mu}\,, && g_{ef_2} = g_{\omega h_1} = \frac{1}{\mu^3}\,, && g_{f_1f_1} = \frac{\alpha}{\mu^3} \,, \\
f_{\omega \omega \omega} = \frac{1}{\mu}\,, && f_{\omega e f_2} = f_{\omega\omega h_1}= f_{ef_1h_1} = \frac{1}{\mu^3}\,, && f_{\omega f_1 f_1} = \frac{\alpha}{\mu^3}\,.
\end{align}
The matrix of Poisson brackets \eqref{Pmat_def} in the flavor space basis $(\omega, e, f_1,h_1,f_2)$ is given by
\begin{equation}\label{PmatECSG}
\cP = \left( \begin{array}{cc}
0 & 0 \\
0 & Q
\end{array} \right)\,,
\end{equation}
with $Q$ given by
\begin{equation}
Q = \frac{1}{\mu^3} \left( \begin{array}{cccc}
\mu^2 V_{ab}^{f_1f_1} - \frac{1}{\alpha} V_{ab}^{h_1h_1} - 2 V_{[ab]}^{f_1f_2} & - \mu^2 V_{ab}^{f_1 e} + V_{ab}^{f_2e} & \frac{1}{\alpha} V_{ab}^{h_1e} & V_{ab}^{f_1e} \\[.2truecm]
- \mu^2 V_{ab}^{ef_1} + V_{ab}^{ef_2} &\mu^2 V_{ab}^{ee} & 0 & - V_{ab}^{ee} \\[.2truecm]
\frac{1}{\alpha} V_{ab}^{eh_1} & 0 & - \frac{1}{\alpha} V_{ab}^{ee} & 0 \\[.2truecm]
V_{ab}^{ef_1} & - V_{ab}^{ee} & 0 & 0
\end{array} \right)\,.
\end{equation}
From \eqref{PmatECSG} it is clear that $\phi_{\rm LL}[\chi]$ defined by \eqref{phiLL} is  first class. To show that the brackets of $\phi_{\rm diff}[\zeta]$ vanish, we may use that, by virtue of
\begin{equation}
e_{[\mu}{}^a f_{1\,\nu]\,a} = e_{[\mu}{}^a h_{1\,\nu]\,a} = e_{[\mu}{}^a f_{2\,\nu]\,a} = 0\,,
\end{equation}
the gauge parameters $\xi^r_a = a_{\mu\,a}^r \zeta^{\mu}$ satisfy
\begin{equation}
e_i{}^a \xi^{f_1}_a = f_{1\,i}{}^a \xi^{e}_{a}\,, \qquad e_i{}^a \xi^{h_1}_a = h_{1\,i}{}^a \xi^{e}_{a}\,, \qquad e_i{}^a \xi^{f_2}_a = f_{2\,i}{}^a \xi^{e}_{a}\,.
\end{equation}
Using these identities, explicit computation shows that $\phi_{\rm diff}[\zeta]$ as defined in \eqref{phidiff} has weakly vanishing brackets with all other primary constraint functions. It is also possible to show that the Poisson brackets of $\phi_{\rm LL}[\chi]$ and $\phi_{\rm diff}[\zeta]$ with the secondary constraints vanish on the AdS vacuum. This is sufficient to identify them as the generators of the gauge symmetries at the AdS boundary, since close to the AdS boundary, we may use the background values for the fields. Then, it becomes possible to split the first class constraint functions into a set of mutually commuting constraints $L_{\pm}$ defined by eq.~\eqref{Lpm}. From the background values of the fields we derive that
\begin{equation}\label{ECSGparam}
\xi^{f_1}_a = \frac{1}{2 \ell^2} \xi^{e}_a\,, \qquad \xi^{h_1}_a = 0 \,, \qquad \xi^{f_2}_a = \frac{1}{2\ell^2}\left( \mu^2 - \frac{\alpha}{4 \ell^2}\right) \xi^{e}_a \,.
\end{equation}
% We define the constraint functions $L_{\pm}[\xi]$ as
% \begin{equation}
% L_{\pm}[\xi] = \phi_{\rm diff}'[\zeta] \pm \frac{1}{\ell} \phi_{\rm LL}[\xi] \,,
% \end{equation}
% where $\phi_{\rm diff}'[\zeta] = \phi_{\rm diff}[\zeta] - \phi_{\rm LL}[\omega_{\mu} \zeta^{\mu}]$.
Upon using the AdS background identities \eqref{ECSGparam} in the expression for the boundary charges \eqref{Qpm}, we find that
\begin{equation}
Q^{\pm}_{3}[\xi^{\pm}] = \pm \frac{1}{8\pi\mu G} \int_{\partial \Sigma} dx^i \; \xi_a^{\pm} \left( \delta \omega_i{}^a \pm \frac{1}{\ell} \delta e_i{}^a \right)\,,
\end{equation}
where we have reinstated the overall factor of $\kappa^2=8 \pi G$. Following the asymptotic analysis of section \ref{sec:cc}, this leads to a central charge given by,
\begin{equation}\label{ECSGcenc}
c_{3} =-\tilde c_{3} = \frac{3}{2\mu G} \,.
\end{equation}
We observe that the result does not depend on the new coupling constant $\alpha$. This is consistent with the result of section \ref{sec:ctheorem} which states that the central charge in the odd sector is universal.

% This could have been
%  anticipated since the $\alpha$-term in the Lagrangian density \eqref{ECSG} vanishes on the AdS background.

% \bibliographystyle{fullsort}
% \bibliography{bimetricbib}

\providecommand{\href}[2]{#2}\begingroup\raggedright\endgroup

\end{document}